\documentclass[aps,prd,twocolumn,superscriptaddress,nofootinbib,%
longbibliography,floatfix]{revtex4-2}

\usepackage{amsmath,amssymb}
\usepackage{graphicx}
\usepackage{bm}
\usepackage[dvipsnames]{xcolor}
\usepackage[colorlinks=true,citecolor=blue,linkcolor=blue,urlcolor=blue]{hyperref}

\newcommand{\Msun}{\,M_\odot}
\newcommand{\kms}{\,{\rm km\,s^{-1}}}
\newcommand{\dd}{d}
\newcommand{\ii}{i}
\newcommand{\ee}{e}
\newcommand{\Rnear}{R_{\rm near}}

\newcommand{\Sigcr}{\Sigma_{\rm cr}}
\newcommand{\deff}{d_{\rm eff}}

\newcommand{\Fgo}{F_{\rm GO}}
\newcommand{\sqrtmu}{\sqrt{\mu}}
\newcommand{\rhosub}{\rho_{\rm sub}}

\newcommand{\unit}[1]{\,\mathrm{#1}}

\begin{document}

\title{Wave-optics imprints of dark matter subhalos on strongly lensed
gravitational waves. II. Saddle images and detectability}

\author{Shin'ichiro Ando}
\affiliation{GRAPPA Institute, University of Amsterdam,
Science Park 904, 1098 XH Amsterdam, The Netherlands}
\affiliation{Kavli Institute for the Physics and Mathematics of the Universe
(WPI), The University of Tokyo, Kashiwa, Chiba 277-8583, Japan}

\date{June 19, 2026}

\begin{abstract}
Wave-optics interference in strongly lensed gravitational waves is a
new interferometric probe of dark matter substructure: a subhalo population
threading a galaxy-scale lens imprints frequency-dependent distortions on the
amplification factor of each macro image. In a companion paper~\cite{Ando:2026poq}, we computed these
imprints for the magnified minimum image.
Here, we extend the calculation to the
saddle-point image and we assess the detectability of the combined signal
with the Laser Interferometer Space Antenna (LISA). Evaluating the amplification
factor at a saddle is numerically delicate, because the equal-arrival-time
contours are open and the subhalo signal is a small difference of large terms; we
present a time-domain method that resolves it. 
Across a Monte Carlo
ensemble of cold dark matter subhalo realizations, subhalos induce percent-level
amplitude and phase modulations in both image parities, while the mean
(de)magnification splits by parity: the minimum is net magnified and the saddle
net demagnified. Demodulating the macro-image interference recovers the per-image
modulations, and a matched-filter analysis that projects out the lens parameters
yields a combined detection above $5\sigma$ in $62\%$ of realizations for
fiducial massive-black-hole-binary sources of total mass $\sim10^{6}\Msun$ at
redshift $1.5$, provided the source lies close to the lens caustic at small impact parameter $y_{\rm src}\lesssim0.1$. Folding these na{\" i}ve per-event significances through optimistic strong-lensing rate
forecasts yields $10$--$20$ substructure detections over the LISA
mission. Strongly lensed gravitational waves are
thus a sensitive, complementary probe of substructure at $10^{4}$--$10^{7}\Msun$
scales inaccessible to electromagnetic observations.
\end{abstract}

\maketitle

\section{Introduction}
\label{sec:intro}

The abundance and density structure of low-mass dark matter subhalos are among the sharpest
discriminants between cold dark matter (CDM) and its
alternatives~\cite{Bullock:2017xww,Zavala:2019gpq}. CDM generically
predicts an abundant population of low-mass, starless subhalos below the threshold
of galaxy formation~\cite{Springel:2008cc}, which warm, self-interacting, and ultralight models do not
reproduce in
detail~\cite{Lovell:2013ola,Dekker:2021scf,Tulin:2017ara,Ando:2024kpk,Ando:2025qtz,
Hui:2016ltb}. The very property that makes these subhalos diagnostic, namely the
absence of stars, places them beyond the reach of most observations. Perturbations
of cold stellar streams~\cite{Banik:2019cza}, flux-ratio anomalies in strongly
lensed quasars~\cite{Vegetti:2023mgp}, and the counts and kinematics of Milky-Way
satellites~\cite{Simon:2019nxf,Nadler:2021dft} each constrain substructure, but
only along a handful of lines of sight, or for the subset of subhalos massive enough to
host baryons. A method that registers the gravity of an individual dark subhalo,
independent of its luminous content, would therefore be a genuinely complementary
probe.

Gravitational waves (GWs) offer such a method. Because a GW is phase coherent, and
because its period can be comparable to a lensing time delay, the lensing of GWs
operates in the wave-optics (WO) regime, in which a perturber leaves a
frequency-dependent imprint on the signal rather than a static
magnification~\cite{Takahashi:2003ix}. Such WO effects have been studied
extensively for compact objects and low-mass perturbers in both ground- and
space-based detectors~\cite{Jung:2017flg,Lai:2018rto,Tambalo:2022wlm,Lin:2023ccz,
Savastano:2023spl,Zumalacarregui:2024ocb,Zumalacarregui:2026uqs,Urrutia:2024pos,Villarrubia-Rojo:2024xcj,Chen:2026qtu}.
Whether that imprint is observable, however,
depends on the geometry. Along generic lines of sight through the cosmological density
field, the effect is weak, and has been forecast to appear in only a small number of
events~\cite{Fairbairn:2022xln,Guo:2022dre,Brando:2024inp,Caliskan:2022hbu,Caliskan:2023zqm}.

In strongly lensed systems, by contrast, it is enhanced for three reasons: (i) the line
of sight pierces the dense inner region of a massive halo; (ii) the macro images
form near critical curves, where small perturbations of the arrival-time surface are
geometrically amplified~\cite{Oguri:2022zpn}; and (iii) the large magnification supplies the high
signal-to-noise ratio (S/N) that the measurement requires. Strong lensing of GWs
occurs at a non-negligible rate for the Laser Interferometer Space Antenna
(LISA)~\cite{Sereno:2010dr,Oguri:2018muv,Gutierrez:2025ymd}, as well as for the
proposed Taiji~\cite{Hu:2017mde,Luo:2021qji} and TianQin~\cite{TianQin:2015yph,Lin:2023ccz} missions, such that multiply imaged
massive-black-hole-binary (MBHB) mergers form a well-defined sample for such a
search.

Much of the WO literature has concentrated on \emph{characterizing}
the lensing imprint (its structure, its phenomenology, and the numerical methods
needed to compute it), while detectability has been assessed mainly for individual perturbers or generic lines of sight. Whether the imprint of a \emph{realistic} dark matter population is a \emph{detectable} probe is a question that the strongly lensed regime makes worth posing and answering. WO signatures of a CDM subhalo population in such
strongly lensed systems have recently begun to be explored, through the WO
diffraction near caustics~\cite{Ezquiaga:2025gkd} and the statistics of higher-order
caustics in the geometric optics (GO) regime~\cite{Vujeva:2025nwg}. In a
companion paper~\cite{Ando:2026poq}, hereafter Paper~I, we made this
program quantitative for the magnified minimum image of a galaxy-scale lens, and
found that a CDM subhalo population imprints percent-level,
frequency-dependent modulations across the LISA band, dominated by subhalos of
$10^{4}$--$10^{7}\Msun$ and measurable in loud MBHB sources.

Paper I, however, left two questions open, which motivate the present paper. The first
concerns image parity. A two-image lens produces a minimum and a saddle, and the two
respond differently to a perturbation. A saddle straddling the critical curve is the
image most readily disrupted by substructure, a fact long exploited in the
flux-ratio-anomaly test, where a perturbed saddle appears anomalously faint relative
to the smooth-lens expectation~\cite{Mao:1997ek,Schechter:2002dm,Dalal:2001fq,Kochanek:2003zc}. In WO, this fragility has so far been
probed only for compact, stellar-mass microlenses embedded in a macro
image~\cite{Diego:2019lcd,Mishra:2021xzz,Yeung:2021chy,Shan:2022xfx}, which likewise
single out the negative-parity saddle as the most strongly distorted image. Whether it
persists for an extended CDM subhalo population, whose members perturb the
image as smooth tidal shear rather than as point masses, is not obvious a priori. Nor is
it obvious in what observable the fragility then resides, since WO records the entire
frequency-dependent transfer function rather than a single flux ratio. The second
question concerns detectability. The per-image amplification factors are not measured
in isolation; the data carry only the total lensed waveform, and a credible estimate
must extract the subhalo signal from that single observable while marginalizing over
the unknown source and lens parameters.

In the present paper, we address both the questions. We compute the WO amplification factor of
the saddle image, for the same lens and the same CDM subhalo realizations as in
Paper~I, and we compare the two parities directly. It is more demanding to 
calculate around the saddle, because its equal-arrival-time contours are open rather than closed,
and the subhalo imprint is a small residual riding on a slowly converging
background. A dedicated time-domain treatment is therefore required, which we
describe and validate against independent references. We then construct a detection
statistic that treats the total waveform as the only observable, demodulating the
macro-image interference to recover the per-image modulations and retaining only the
part of the subhalo signal that cannot be reabsorbed into the source and lens
parameters. We find that the
saddle is as informative as the minimum, and that the GO fragility
survives into WO, not as a larger fluctuation, but as a systematic parity
split in the mean (de)magnification.

The paper is organized as follows. Section~\ref{sec:model} summarizes the lens
model, the subhalo population, and the WO amplification factor common to
both image parities. Section~\ref{sec:saddle} presents the saddle-point calculation
and its validation, and Section~\ref{sec:detect} develops the detectability
formalism. Section~\ref{sec:results} reports the per-image modulations, the parity
asymmetry, and the LISA detection significances over the Monte Carlo ensemble,
Section~\ref{sec:discussion} discusses the origin and diagnostic value of the
saddle signal, and Section~\ref{sec:conclusions} concludes. Technical details are
collected in the Appendices. We adopt the \textit{Planck}~2018
cosmology~\cite{Planck:2018vyg} throughout.

\section{Model and formalism}
\label{sec:model}

Our lens model, subhalo population, and WO formalism follow Paper~I,
to which we refer for full derivations. We summarize them here to fix notation and
to emphasize the elements that the saddle-point analysis requires. The
construction is a two-level hierarchy: a smooth \emph{macrolens} that produces the
strongly lensed images, and a population of low-mass \emph{subhalos} whose
WO imprint on each image we wish to compute.

\subsection{Macrolens and image configuration}
\label{sec:macro}

The macrolens is a host dark matter halo with a central galaxy and its heaviest
subhalos. The host has a Navarro--Frenk--White (NFW) density
profile~\cite{Navarro:1995iw,Navarro:1996gj} of virial mass
$M_{200}=10^{12}\Msun$ at lens redshift $z_L=0.5$, with a concentration parameter
$c_{200}=10/(1+z_L)$ from the mass--concentration--redshift relation
of Ref.~\cite{Correa:2015dva}. The scale radius and characteristic density follow from
\begin{equation}
\begin{gathered}
R_{200}=\left(\frac{3M_{200}}{4\pi\cdot 200\rho_{\rm c}(z_L)}\right)^{1/3},\quad
R_{\rm s}=\frac{R_{200}}{c_{200}},\\
\rho_{\rm s}=\frac{M_{200}}{4\pi R_{\rm s}^{3}\,\nu(c_{200})},
\end{gathered}
\end{equation}
with $\nu(c)=\ln(1+c)-c/(1+c)$. The central galaxy is modeled as a singular
isothermal sphere (SIS) with velocity dispersion $\sigma_v=250\kms$, and we
include the heaviest subhalos with masses $m>10^{-3}M_{200}=10^{9}\Msun$, drawn from the
semi-analytic \textsc{SASHIMI}
model\footnote{\textsc{SASHIMI}: \url{https://github.com/shinichiroando/sashimi-c}.}~\cite{Hiroshima:2018kfv,Ando:2019xlm,
Ando:2020yyk} and distributed in three dimensions according to the host number
density $n_{\rm sub}(r)\propto(r^{2}+R_{\rm s}^{2})^{-3/2}$. Lengths are made
dimensionless with $\xi_0=R_{\rm s}$, and the critical surface density is
\begin{equation}
\Sigcr=\frac{c^{2}}{4\pi G(1+z_L)\,\deff},~~
\deff=\frac{D_L D_{LS}}{(1+z_L)D_S},
\end{equation}
with $D_{L}$, $D_S$, and $D_{LS}$ the angular-diameter distances to lens, source, and between
them. The source is at $z_S=1.5$ with dimensionless impact parameter
$y_{\rm src}=0.1$ in units of the SIS Einstein radius for our fiducial
configuration.

The images are the stationary points of the Fermat (arrival-time) potential
\begin{equation}
\phi(\bm{x},\bm{y})=\frac{1}{2}\,|\bm{x}-\bm{y}|^{2}-\psi_{\rm macro}(\bm{x}),
\label{eq:fermat}
\end{equation}
$\nabla\phi=0$, 
where $\psi_{\rm macro}$ is the projected potential of the host, galaxy, and heavy
subhalos, and $\bm{y}$ is the source position. For our fiducial configuration the
macrolens produces the generic two-image system of a non-singular lens: a
\emph{minimum} (Type~I, $\det A>0$, $\mathrm{tr}\,A>0$) and a \emph{saddle}
(Type~II, $\det A<0$), where
\begin{equation}
A(\bm{x})=\frac{\partial^{2}\phi}{\partial\bm{x}\,\partial\bm{x}}
=I-\nabla\nabla\psi_{\rm macro}(\bm{x})
\end{equation}
is the lens-mapping Jacobian and $\mu=1/|\det A|$ the magnification. In the
LISA band, the macro-image time delay (tens of days for a galaxy lens) is far
larger than the GW period. The interference \emph{between} macro images is therefore in
the GO limit. Any measurable frequency dependence of a single
image must therefore originate in the WO response of its local subhalo
population. While Paper~I analyzed the minimum alone, here we treat the minimum and the saddle
on the same footing.

\subsection{Subhalo population and selection}
\label{sec:subs}

Around each macro image, we populate low-mass subhalos in the range
$10^{2}$--$10^{9}\Msun$ ($10^{-10}$--$10^{-3}$ in units of $M_{200}$), again drawn
from \textsc{SASHIMI}. Each subhalo's density is characterized by a truncated NFW
profile~\cite{Baltz:2007vq},
\begin{equation}
\rho(r)=\frac{\rho_{\rm s}}{(r/r_{\rm s})(1+r/r_{\rm s})^{2}}
\left(\frac{r_{\rm t}^{2}}{r_{\rm t}^{2}+r^{2}}\right),
\label{eq:tnfw}
\end{equation}
whose finite total mass yields a single clean logarithmic far field for $r\gg r_{\rm t}$, a property we
will exploit in the saddle calculation (Sec.~\ref{sec:saddle}). Because
\textsc{SASHIMI} fixes the truncation radius assuming an abrupt three-dimensional
cut, we remap the truncation of the smooth profile [Eq.~(\ref{eq:tnfw})] such that its enclosed mass
equals the subhalo bound mass exactly. We implement this profile directly as a
native lens potential in the public WO solver
\textsc{GLoW}\footnote{\textsc{GLoW}: \url{https://github.com/miguelzuma/GLoW_public}.}~\cite{Tambalo:2022plm,Villarrubia-Rojo:2024xcj},
by extending its built-in lens library rather than approximating the truncated halo
with a combination of existing components; the mass remap and the closed-form
truncated-NFW lensing functions are given in Appendix~\ref{app:population}.

A subhalo contributes to the WO integral only if it perturbs the image on
scales the GW can resolve. We classify each subhalo by two dimensionless
frequencies evaluated at the low end of the band, $f_{\min}=10^{-4}\unit{Hz}$,
\begin{equation}
w_{\rm sub}=\frac{2\pi f_{\min}\,r_{\rm s}^{2}}{c\,\deff},~~
w_{E}=\frac{4\pi G\,m\,f_{\min}}{c^{3}},
\end{equation}
the former set by the subhalo scale radius and the latter by its Einstein time
delay. A subhalo is treated in the GO limit when either frequency exceeds a threshold
$w_{\rm th}=10^{2}$.
The GO subhalos are folded into the macro lens, and then, the macro
images and their Jacobian are re-evaluated to include them
(Appendix~\ref{app:population}). Only the remaining WO subhalos
enter the diffraction integral. As in Paper~I, the
signal is dominated by $10^{4}$--$10^{7}\Msun$ subhalos, whose Einstein time delays
$\sim 4Gm/c^{3}$ overlap the LISA band.

Subhalos are sampled within a projected radius $\Rnear$ of the
image, with surface density held fixed.
The perturbation radius~\cite{Ando:2026poq}
\begin{equation}
\Rnear=\max\!\big[N_F R_F,\ R_\mu,\ R_{\rm core}\big],~~
R_\mu=\sqrt{\frac{g_{\rm img}\,m}{\pi\,\epsilon_\mu\,\Sigcr}},
\label{eq:rnear}
\end{equation}
combines the Fresnel scale $R_F=[c\,\deff/(2\pi f_{\min})]^{1/2}$ (with $N_F=5$),
the magnification-perturbation radius $R_\mu$ at which a subhalo changes the GO
magnification by a fraction $\epsilon_\mu=10^{-2}$, and a core floor
$R_{\rm core}=\min[N_E r_{\rm s},r_{\rm t}]$ with $N_E=5$. Here
$g_{\rm img}=\lVert A^{-1}\rVert$ is the local magnification amplification through
the inverse Jacobian, which is larger near the critical curve and is the geometric
origin of the enhanced response of strongly lensed images. 

\subsection{Per-image wave-optics amplification factor}
\label{sec:wo}

For a single macro image, we work in a local WO frame centered on that
image, with dimensionless frequency $w=f/f_0$ ($f_0=3\unit{mHz}$ in our
calculations); the length unit $\xi_0$ is redefined accordingly, from the macro value
$R_{\rm s}$ to the Fresnel scale at $f_0$, $\xi_0=[c\,\deff/(2\pi f_0)]^{1/2}$.
The amplification factor is the Fresnel--Kirchhoff diffraction
integral~\cite{Nakamura:1999uwi,Takahashi:2003ix} (see also
Ref.~\cite{CarrilloGonzalez:2025gqm} for recent theoretical developments)
\begin{equation}
F(w)=\frac{w}{2\pi\ii}\int \dd^{2}u\,
\exp\!\left[\ii w\left(\frac{1}{2}\bm{u}^{\mathsf T}\!A\,\bm{u}
-\delta\psi(\bm{u})\right)\right],
\label{eq:Fw}
\end{equation}
where $\bm{u}=\bm{x}-\bm{x}_{\rm img}$ is the offset from the image, $A$ is the
macro Jacobian at the image, and $\delta\psi$ is the projected potential of the
local WO subhalos. The quadratic macro term is exact to second order
about the stationary point, which guarantees that the GO image is preserved and
that the macro field is not double-counted with the explicit subhalo term. The
external-field decomposition is summarized in Appendix~\ref{app:population} and
follows Paper~I.

It is convenient to compute $F(w)$ in the time domain.\footnote{The Fresnel--Kirchhoff
integral can alternatively be evaluated directly in the frequency domain, for which
fast Fourier-transform (FFT) methods have recently been developed~\cite{Ephremidze:2026era}.}
Inserting the identity
\begin{equation*}
1=\int\dd\tau\,\delta(\tau-\phi)
\end{equation*}
into Eq.~\eqref{eq:Fw} gives
\begin{equation}
F(w)=\frac{w}{2\pi\ii}\int \dd\tau\,I(\tau)\,\ee^{\ii w\tau},~~
I(\tau)=\oint_{\phi=\tau}\frac{\dd\ell}{|\nabla\phi|},
\label{eq:Itau}
\end{equation}
where $I(\tau)$ is the integral over the equal-arrival-time contour
$\{\phi=\tau\}$, equivalently the rate of growth of the area enclosed by that
contour, $I(\tau)=\dd A/\dd\tau$ (the co-area identity, evaluated by
contour following as in Refs.~\cite{Ulmer:1994ij,Villarrubia-Rojo:2024xcj}). For a pure quadratic
image ($\delta\psi=0$), the GO limit $F(w)\to\Fgo$ is recovered: a minimum has the
constant $I_{\rm GO}=2\pi/\sqrt{|\det A|}=2\pi\sqrtmu$ and $\Fgo=\sqrtmu$, while a
saddle has an open, logarithmically divergent $I(\tau)$ [Eq.~\eqref{eq:Iquad}]
whose transform gives $\Fgo=-\ii\sqrtmu$, the factor $-\ii$ being the Morse phase
$\ee^{-\ii\pi n/2}$ with Type-II index $n=1$~\cite{Dai:2017huk}. Subhalos distort the \emph{shape} of
$I(\tau)$ while leaving its normalization nearly fixed, and these shape distortions
are what produce the frequency-dependent modulation of $F(w)$. The central object
of this paper is the per-image \emph{envelope},
\begin{equation}
\eta(w)\equiv \frac{F(w)}{\Fgo}-1,
\label{eq:mdef}
\end{equation}
the complex modulation of the full lens (macro $+$ subhalos) relative to the
subhalo-free GO reference: $|1+\eta|$ measures the amplitude distortion and $\arg(1+\eta)$
the phase distortion imprinted by the local subhalo population on that image. For
the minimum, $I(\tau)$ is a closed loop and $F(w)$ is straightforward to evaluate,
as in Paper~I. For the saddle, on the other hand, evaluating $I(\tau)$ involves an open contour and 
requires the dedicated treatment of the next section.

\section{The saddle-point calculation}
\label{sec:saddle}

\begin{figure*}[t]
\centering
\includegraphics[width=0.92\textwidth]{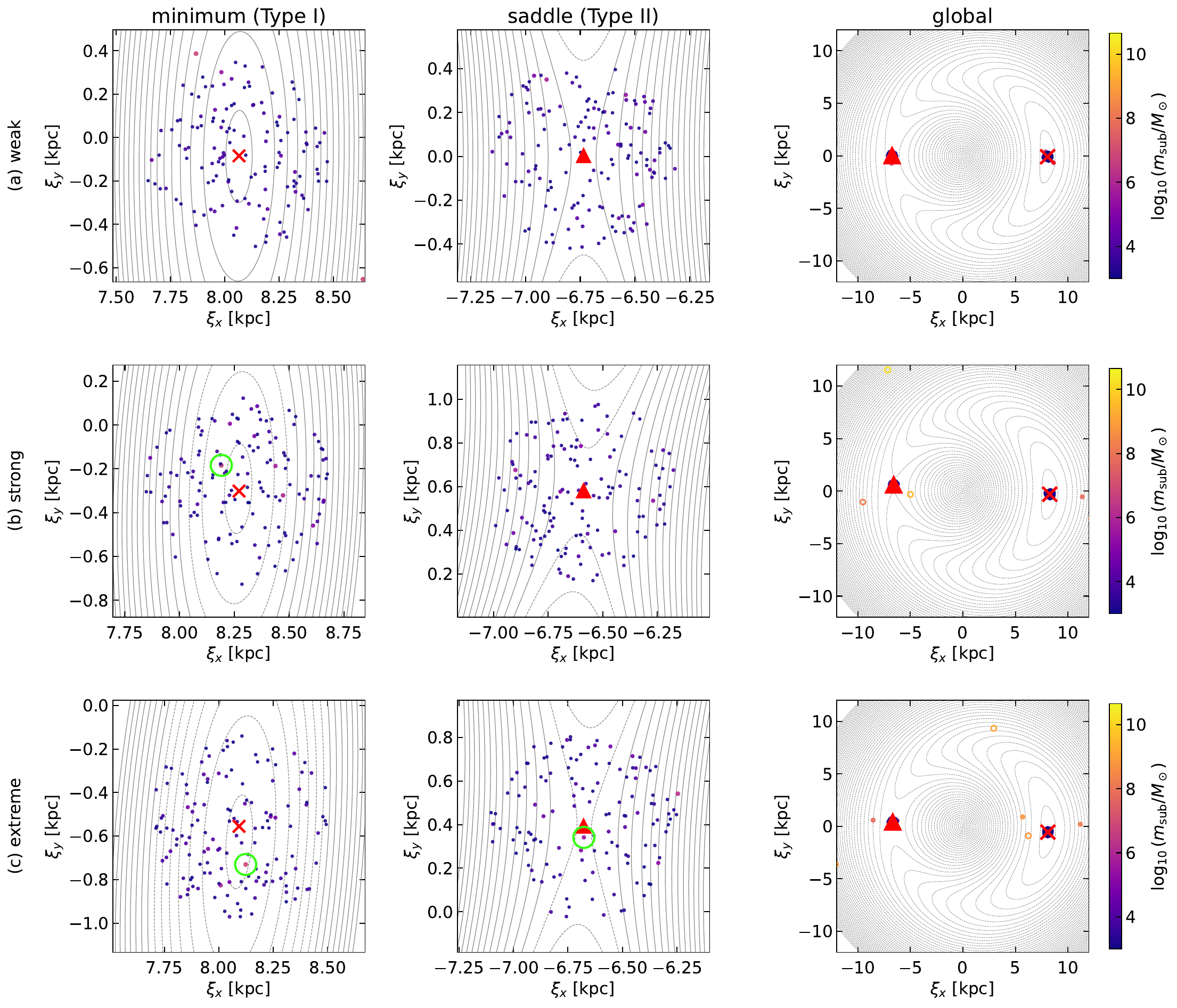}
\caption{Iso-arrival-time (iso-Fermat) topology for three realizations, one per row,
selected to span the range of subhalo detectability: (a) weak (sub-threshold),
(b) strong, and (c) extreme cases (Sec.~\ref{sec:res_detect}). These rows are illustrative
examples and are \emph{not} drawn at random from the ensemble. The columns are the
minimum (Type~I, red cross; left) and saddle (Type~II, red triangle; center) local panels and the global macro
panels (right).
All panels use the physical lens-plane
coordinate $\xi_{x,y}$ in kpc, with the halo center at the origin.  WO (GO) subhalos are shown as filled (open) circles, colored by mass. For legibility, the scatter points show only the subhalos with $m_{\rm sub}>10^{3}\Msun$, while $F(w)$ calculations also include the more numerous lighter perturbers down to $10^{2}\Msun$.
The green circle in each local panel marks a single subhalo whose removal lowers that
image's detection significance by more than $50\%$ (a leave-one-out test,
Sec.~\ref{sec:res_detect}). Such a dominant perturber is present only in the strong and
extreme minimum and the extreme saddle, in each case a $\sim10^{6}\Msun$ perturber within
$b\lesssim0.01\,R_{\rm s}$ of the image.
}
\label{fig:topology}
\end{figure*}

The saddle image is both the physically interesting case and the numerically
demanding one. We first explain why, then describe the method we use to evaluate
its amplification factor, and finally validate it. Full technical details are
deferred to Appendix~\ref{app:saddle}. Figure~\ref{fig:topology} previews the
distinction at the root of the difficulty: the minimum's closed equal-arrival-time
contours against the saddle's open, non-closing ones.

\subsection{Why the saddle is hard}
\label{sec:saddle_hard}

At a saddle, the Jacobian $A$ has eigenvalues of opposite sign, $\lambda_+>0>
\lambda_-$, and the magnification is $\sqrtmu=1/\sqrt{\lambda_+|\lambda_-|}$.
Writing the local arrival-time surface in eigencoordinates,
$\phi-\phi_{\rm sad}\simeq\lambda_+ x_+^{2}/2-|\lambda_-|x_-^{2}/2$,
the equal-arrival-time contour $\{\phi=\tau\}$ is not a closed loop, but a pair of \emph{open}, hyperbola-like arcs that run outward to the
boundary of the region in which the quadratic approximation holds. The co-area
integral of Eq.~\eqref{eq:Itau} for the bare quadratic saddle is
\begin{equation}
I_{\rm quad}(s)=\frac{4}{\sqrt{\lambda_+|\lambda_-|}}\,
\mathrm{arccosh}\!\left(\sqrt{\frac{\lambda_+\big(|\lambda_-|R_{\rm c}^{2}+2|s|\big)}
{2|s|\,(\lambda_++|\lambda_-|)}}\,\right),
\label{eq:Iquad}
\end{equation}
where $s\equiv\tau-\phi_{\rm sad}$, $R_{\rm c}$ is the radius of the circular
patch $x_+^2+x_-^2\le R_{\rm c}^2$ at which the open contour is cut, and the two
hyperbola branches (each summed through both vertex halves) contribute. The
expression is written for $s>0$, with the $s<0$ branch following from
$\lambda_+\leftrightarrow|\lambda_-|$. This patch-bounded closed form is the
finite-region saddle template of Ref.~\cite{Tambalo:2022plm}, with their
contour limit $\delta\tau$ here resolved into the explicit patch geometry. We
derive it from the co-area integral~\eqref{eq:Itau} in
Appendix~\ref{app:quadtemplate}. Unlike the minimum's GO
constant $I_{\rm GO}=2\pi\sqrtmu$~\cite{Tambalo:2022plm,Villarrubia-Rojo:2024xcj},
this depends on the radius $R_{\rm c}$ at which the contour leaves the quadratic
patch and grows logarithmically with the patch size. 

The subhalo imprint we want is the
\emph{difference}
\begin{equation}
\delta I(s)\equiv I_{\rm full}(s)-I_{\rm quad}(s)
\end{equation}
between the full (macro $+$ WO subhalo) contour integral and this quadratic GO template. Three
features make recovering it numerically delicate, none of which troubles the
minimum.

\begin{enumerate}
\item The iso-arrival contour does not close. As Eq.~\eqref{eq:Iquad} shows, the
open-arc template is not a finite closed-loop value but diverges, growing
logarithmically with the patch radius $R_{\rm c}$ at which the contour is cut.

\item The subhalo signal rides on a divergent background. Toward the saddle delay
($s\to0$), the template $I_{\rm quad}$ diverges as $-2\sqrtmu\ln|s|$, and $\delta I$
is a percent-level residual sitting on top of this large, singular background.
Recovering it requires subtracting the macro background accurately where it is
largest, rather than reading the signal off directly as one can at the minimum,
whose background is finite.

\item The perturbing potential reaches to infinity. An NFW subhalo falls off too
slowly for its imprint to be compact, leaving a non-cancelling tail in $\delta I$
that cannot be traced to arbitrarily large delay. We therefore adopt the truncated-NFW profile of Eq.~\eqref{eq:tnfw}: its finite total mass gives a single clean
far-field logarithm $\psi\to m\ln r$ that can be subtracted analytically
(Sec.~\ref{sec:tail}), whereas an untruncated or compensated profile would leave a
non-cancelling tail in $I(s)$.
\end{enumerate}

These three are why Paper~I deferred the saddle, and why the public \textsc{GLoW}
solver~\cite{Villarrubia-Rojo:2024xcj}, whose time-domain engine assembles a single
closed arrival-time contour, evaluates the minimum directly but does not on its own
yield the saddle's open, multi-branch $I(\tau)$ or its transform. We evaluate it
with the dedicated method below.

\subsection{Method}
\label{sec:saddle_method}

The saddle and minimum calculations of Paper~I and the present work are carried
out in a single code, \textsc{Sazanami}.\footnote{\textsc{Sazanami}: Strong-lensing
Amplification at $Z$, Arrival-time Numerics, and Modulation Imprints. The repository
will be made publicly available shortly.} We evaluate the saddle
amplification factor in the time domain by
separating an analytic macro GO template from a numerically transformed WO subhalo residual,
\begin{equation}
F(w)=\ee^{\ii w\phi_{\rm sad}}\!\left[\,-\ii\sqrtmu
+\frac{w}{2\pi\ii}\Big(\mathrm{Filon}\!\big[\delta I\big](w)+\widetilde{T}(w)\Big)\right],
\label{eq:recipe}
\end{equation}
where the first term $\Fgo = -\ii\sqrtmu$ is the GO amplitude of the bare
quadratic saddle. It is the \emph{exact} transform of the patch-bounded
template $I_{\rm quad}$, the Fresnel integral of a quadratic arrival-time surface
being closed-form at all $w$. The subhalo imprint at low to moderate delay is carried by
$\mathrm{Filon}[\delta I]$, a Filon quadrature for highly oscillatory
integrals~\cite{Iserles:2005} of the traced, compactly supported residual
$\delta I(s)=I_{\rm full}(s)-I_{\rm quad}(s)$. It integrates the oscillatory
kernel $\ee^{\ii ws}$ analytically on each panel, staying accurate across the
band on the coarse, non-uniform $s$-grid clustered toward $s\to0$ on which
$\delta I$ is sampled, without the fine uniform resampling a plain FFT would
demand. The Filon quadrature carries the transform out to a matching delay,
$|s|<s_{\rm hi}=6000$,
beyond which the analytic far-field tail $\widetilde{T}$ takes over
(Sec.~\ref{sec:tail}).
This residual is smooth at $s\to0$ because the GO logarithm
common to $I_{\rm full}$ and $I_{\rm quad}$ cancels in the difference, making the
quadrature well behaved. The same difference also removes the dependence on the
patch radius: the
$R_{\rm c}$-divergent macro part common to $I_{\rm full}$ and $I_{\rm quad}$
cancels in $\delta I$.
Finally, $\widetilde{T}(w)$ is the analytic transform of
the far-field tail, the part of the residual at large delay where, far from the
subhalo cores, $\delta I$ has settled onto a closed-form $(a+b\ln s)/s$ asymptote
fixed by the truncated halo monopole and is continued analytically (Sec.~\ref{sec:tail}).

The numerical core of the recipe is the traced $I_{\rm full}(s)$. We evaluate it
by following the two open, hyperbola-like iso-arrival-time arcs of the full lens
outward from the saddle to the patch radius $R_{\rm c}$ and computing the co-area
integral~\eqref{eq:Itau} along them, with the subhalo deflections entering
$\nabla\phi$ pointwise such that their imprint enters $I_{\rm full}$ directly. The
arcs are followed with an adaptive, error-controlled open-contour integrator,
detailed in Appendix~\ref{app:trace}. We
validate the traced $I(s)$ against an independent, contour-free area-method
arbiter, which it reproduces to $\sim10^{-3}$.

\subsection{The saddle far-field tail}
\label{sec:tail}

Beyond the Filon window, the residual has settled onto a closed-form asymptote that
we continue analytically. This is the $\widetilde{T}(w)$ of
Eq.~\eqref{eq:recipe}. Per Fermat-delay branch ($s\gtrless0$),
\begin{equation}
\begin{aligned}
\widetilde{T}(w)&=C_+J(x)+C_-\overline{J}(x)\\
&\quad+D_+K(x)+D_-\overline{K}(x),\qquad x=w\,s_{\rm hi},
\end{aligned}
\label{eq:Ttilde}
\end{equation}
with the cosine/sine-integral and logarithmic kernels
\begin{equation}
\begin{aligned}
J(x)&=-\mathrm{Ci}(x)+\ii\left[\frac{\pi}{2}-\mathrm{Si}(x)\right],\\
K(x)&=\int_{s_{\rm hi}}^{\infty}\frac{\ln s}{s}\,\ee^{\ii ws}\,\dd s,
\end{aligned}
\label{eq:JK}
\end{equation}
and the per-branch amplitudes
\begin{equation}
\begin{aligned}
C_\pm&=\sqrtmu\,M_{\rm enc}^{\pm}(\ln2+\gamma),\\
D_\pm&=\sqrtmu\,M_{\rm enc}^{\pm}.
\end{aligned}
\label{eq:tailC}
\end{equation}
The two factors separate cleanly. The amplitudes $C_\pm,D_\pm$ are fixed by the
lens. The enclosed subhalo monopole $M_{\rm enc}^{\pm}$ (the superscript
$\pm$ labels the two Fermat-delay branches $s\gtrless0$, whose open arcs run outward
through different regions of the lens plane and therefore enclose different subhalo
monopoles) and the saddle,
through $\sqrtmu=1/\sqrt{\lambda_+|\lambda_-|}$ and
$\gamma=\ln[(\lambda_+^{-1}+|\lambda_-|^{-1})/4]$.
The kernels $J,K$ depend only on $x=w\,s_{\rm hi}$, with $s_{\rm hi}=6000$ the
Filon-to-tail matching delay introduced in Sec.~\ref{sec:saddle_method}.
Derivations are given in Appendix~\ref{app:tail}.

Physically, $C_\pm,D_\pm$ are the per-branch coefficients of the asymptotic
residual left by the truncated halo monopole $\psi\simeq M_{\rm t}\ln r$,
\begin{equation}
\delta I(s)\xrightarrow[s\to\infty]{}\frac{a+b\ln s}{s},\qquad
a,\,b\propto\sqrtmu\,M_{\rm t},
\label{eq:tail}
\end{equation}
with $M_{\rm t}=m/(\pi\xi_0^2\Sigcr)$ a dimensionless subhalo mass.
For a population, the relevant monopole is the mass $M_{\rm enc}$ \emph{enclosed
within the window} ($|s_i|<s_{\rm hi}$). 
The $\ln s/s$ enhancement is the
mathematical signature distinguishing a saddle from a minimum: the logarithmic
potential is sampled over the open contour's logarithmically growing extent
$t_{\max}\sim\ln s$ [Eq.~\eqref{eq:Iquad}], the product of two logarithms; a
minimum, whose contour is closed and bounded, lacks the second and its residual
falls as a pure $1/s$.

Including $\widetilde{T}$ makes the envelope independent of the matching point
$s_{\rm hi}$, but its imprint is small beside the Filon interior it supplements.
Across our realizations, $\widetilde{T}$ 
contributes at most a few percent of the Filon-driven modulation (median $0.2\%$), which leaves the detection
statistics unchanged. We nevertheless retain the term for
completeness.

\section{Wave-optics modulations and the parity asymmetry}
\label{sec:results}

We generate a Monte Carlo ensemble of $1000$ realizations, sampling the
light subhalos within $\Rnear$ of each image (Sec.~\ref{sec:subs}). Each
realization fixes one macrolens and one WO subhalo catalog drawn down to
$100\Msun$. For each, we compute the per-image amplification factor $F(w)$ and
modulation $\eta(w)$ for both the minimum and the saddle, over the LISA frequency band
$f\in[10^{-4},10^{-1}]\unit{Hz}$. The same
realizations are reused for both image parities so that the comparison between
them is paired and the macro magnification cancels. Unless stated otherwise, the
results below are for the fiducial $y_{\rm src}=0.1$ ensemble.

\subsection{Per-image wave-optics modulations}
\label{sec:res_mod}

\begin{figure}[t]
\centering
\includegraphics[width=0.95\columnwidth]{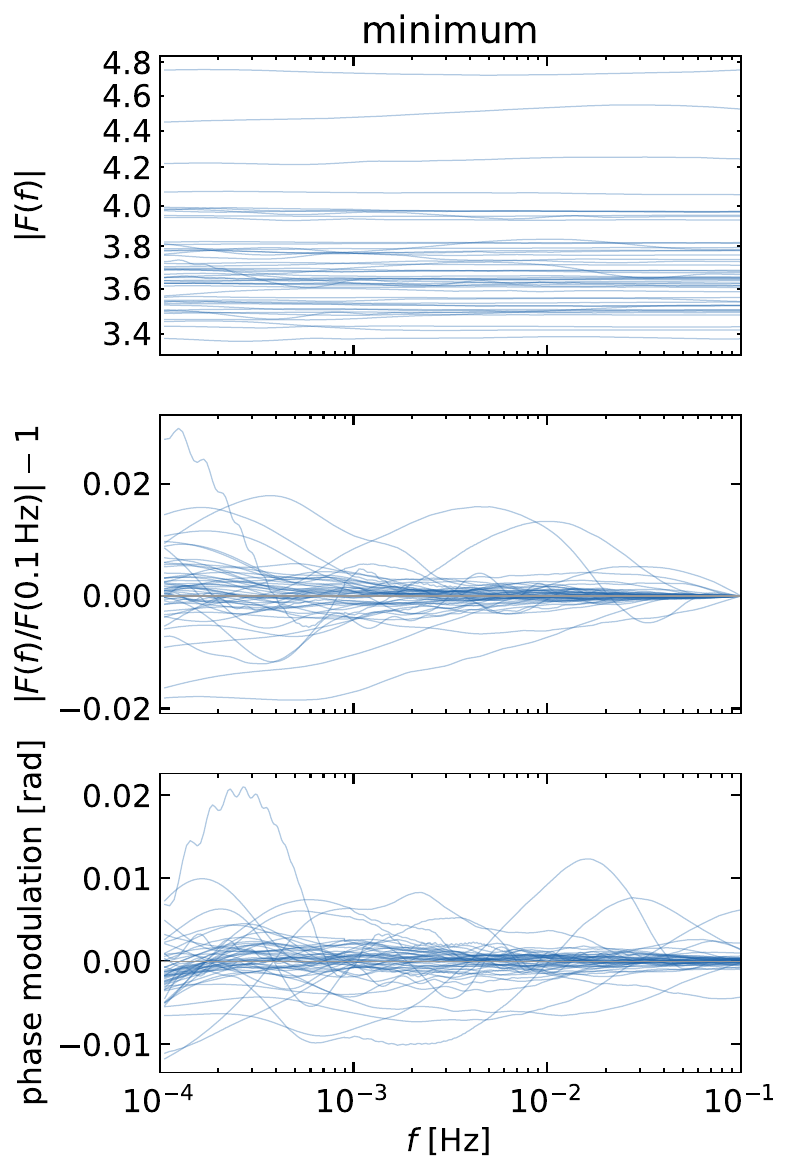}
\caption{WO signal of the \emph{minimum} image versus frequency, shown
for $50$ randomly sampled realizations: the amplification factor
$|F(f)|$ (top), the relative amplitude modulation $|F/F(0.1\,\mathrm{Hz})|-1$
(middle), and the phase modulation (bottom).}
\label{fig:mod_min}
\end{figure}

\begin{figure}[t]
\centering
\includegraphics[width=0.95\columnwidth]{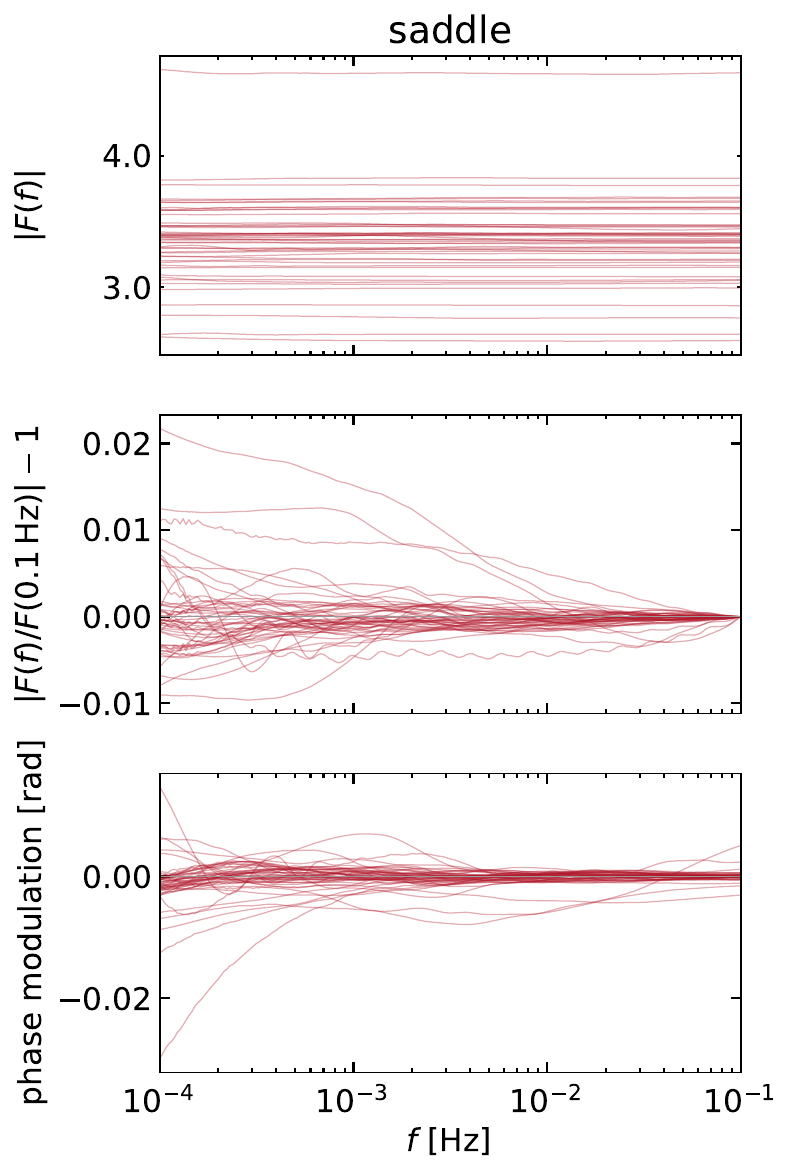}
\caption{The same as Fig.~\ref{fig:mod_min}, but for the \emph{saddle} image. The fluctuation
amplitude is comparable to the minimum (parity-blind), while the mean offset
differs.}
\label{fig:mod_sad}
\end{figure}

Figures~\ref{fig:mod_min} and~\ref{fig:mod_sad} show the per-image modulation as a
function of frequency for the minimum and the saddle, with $50$ randomly sampled
realizations per parity drawn as semi-transparent lines: the amplification factor
$|F(f)|$ (top), the relative amplitude modulation $|F/F(0.1\,\mathrm{Hz})|-1$
(middle), and the phase modulation (bottom).
As in Paper~I for the minimum, the median amplification
closely tracks the GO magnification, but the ensemble shows a
non-negligible dispersion: subhalos imprint amplitude modulations at the percent
level, largest at the low-frequency end $f\lesssim10^{-3}\unit{Hz}$, accompanied by
phase shifts of order $10^{-2}\unit{rad}$. 

\begin{figure}[t]
\centering
\includegraphics[width=0.95\columnwidth]{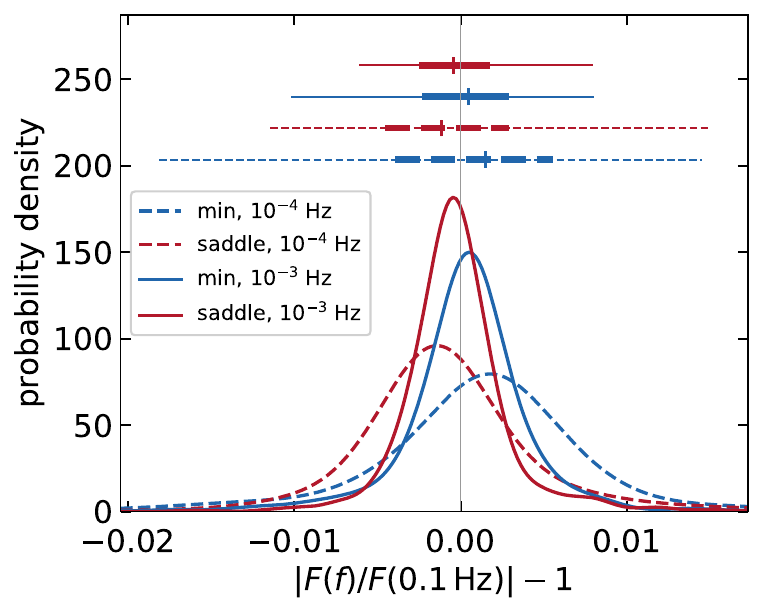}
\caption{Distribution function of the amplitude modulation across 1000 realizations at
$f=10^{-4}\unit{Hz}$ (dashed) and $10^{-3}\unit{Hz}$ (solid), for the minimum (blue)
and saddle (red). The bars above the curves mark the $68\%$ (thick) and $95\%$ (thin)
intervals, with a tick at the median. 
The three example realizations (a), (b), and (c) of Fig.~\ref{fig:topology} lie at the
$28$th, $92$nd, and $99.7$th percentiles of the amplitude-modulation amplitude.}
\label{fig:dist}
\end{figure}

\begin{figure}[t]
\centering
\includegraphics[width=0.95\columnwidth]{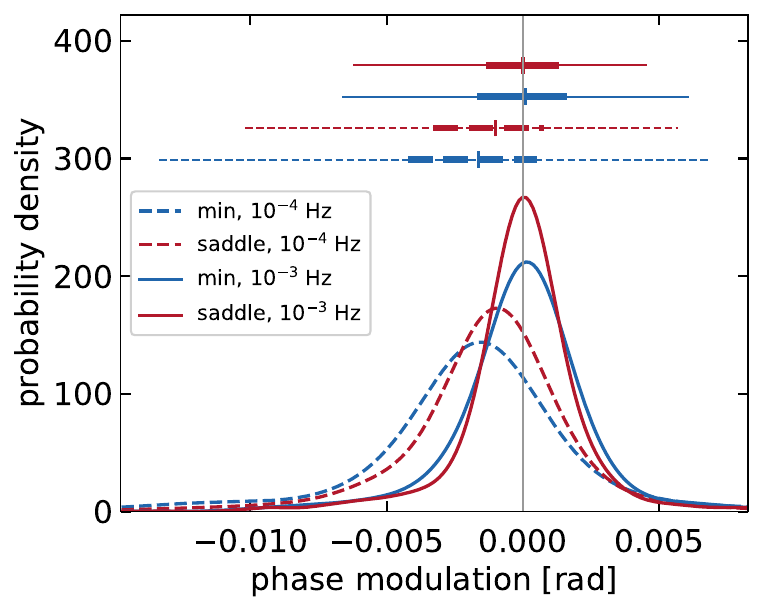}
\caption{The same as Fig.~\ref{fig:dist}, but for the \emph{phase} modulation $\arg(1+\eta)$.  The (a), (b),
and (c) realizations of Fig.~\ref{fig:topology} lie at the $46$th, $96$th, and
$99.7$th percentiles of the phase-modulation amplitude.}
\label{fig:dist_phase}
\end{figure}

The saddle behaves analogously, with
comparable modulation amplitude: the \emph{fluctuation} amplitude of the subhalo
imprint is essentially parity-blind, the WO counterpart of the well-known
result that substructure perturbs both image types at a similar fractional level.
Figures~\ref{fig:dist} and \ref{fig:dist_phase} make this explicit, showing the 
distribution of the amplitude and phase modulations, respectively, across 1000 realizations at two representative frequencies,
$f=10^{-4}$ and $10^{-3}\unit{Hz}$, for the two parities. The distributions are
broad and similar for the minimum and the saddle.

\subsection{The parity (de)magnification asymmetry}
\label{sec:res_parity}

\begin{figure}[t]
\centering
\includegraphics[width=0.95\columnwidth]{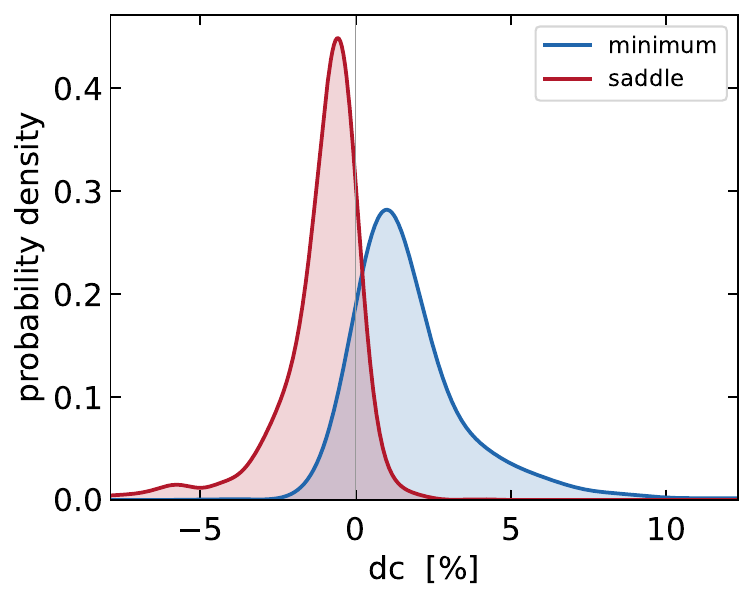}
\caption{Ensemble distribution of the net (de)magnification $\mathrm{dc}$ for the
two parities, oppositely skewed: a magnification tail for the minimum and a
demagnification tail for the saddle.
}
\label{fig:parity_skew}
\end{figure}

Although the fluctuation amplitude is parity-blind, the \emph{mean} of the
modulation is not. For each image, we reduce the modulation to its band-averaged
direct-current (dc) component,
\begin{equation}
\mathrm{dc}\equiv\left\langle \left|\frac{F(f)}{F_{\rm GO}}\right|-1 \right\rangle_f,
\end{equation} the net WO
(de)magnification relative to the GO-only macro reference $|F_{\rm GO}| = \sqrt{\mu}$. 
Figure~\ref{fig:parity_skew} shows the ensemble distribution of $\mathrm{dc}$ for the
two parities.
The two distributions are correspondingly skewed in opposite senses, the minimum with a
positive (magnification) tail and the saddle with a negative (demagnification) tail.
The median of the distributions is $\mathrm{dc}\simeq+1.3\%$ for the minimum and 
$\mathrm{dc}\simeq-0.8\%$ for the saddle.
Paired by realization, the saddle lies systematically
below its own minimum in the vast majority of realizations.

\begin{figure*}[t]
\centering
\includegraphics[width=0.86\textwidth]{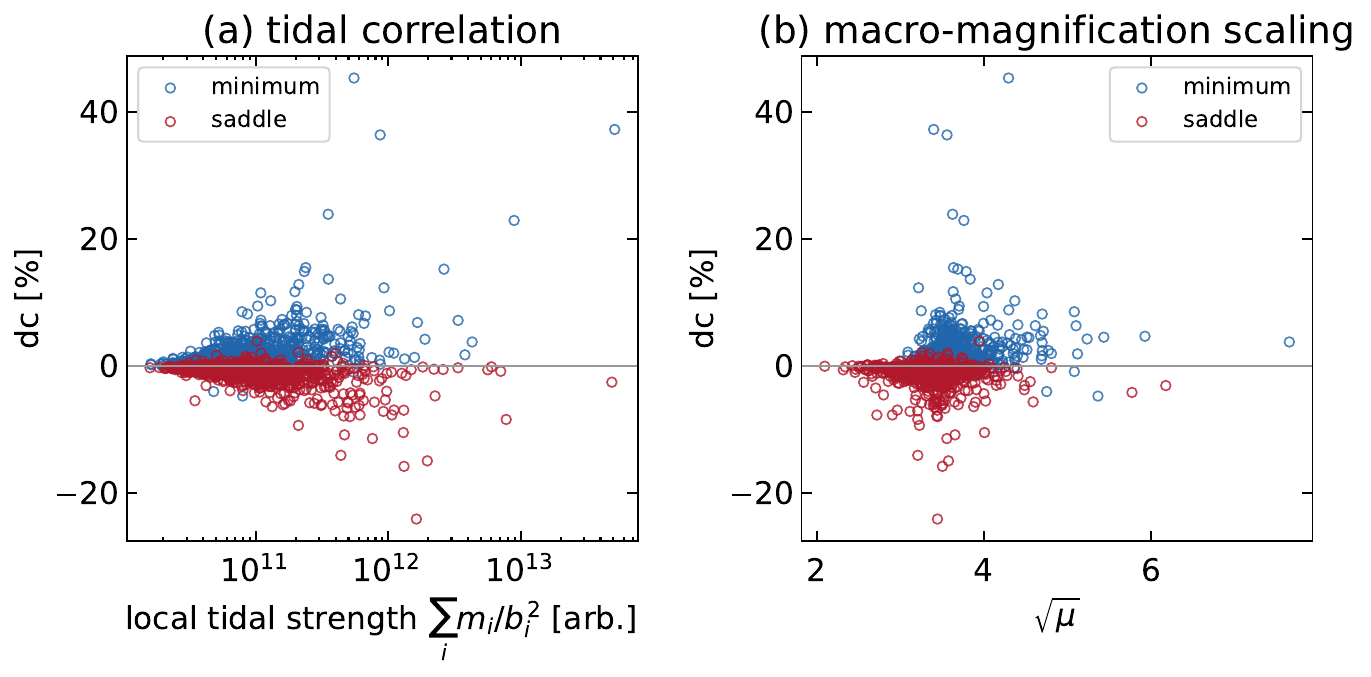}
\caption{Physical origin of the parity asymmetry. (a) $\mathrm{dc}$ versus the local tidal strength
$T=\sum_i m_i/b_i^2$, correlated with \emph{opposite sign} for the minimum (blue,
$+$) and the saddle (red, $-$); the opposite-sign correlation with a physical
quantity ($T$) shows the parity (de)magnification is a real effect, not a numerical
artifact.
(b) $\mathrm{dc}$ versus the macro
magnification $\sqrtmu$. The magnitude grows toward the critical curve, i.e., with
increasing magnification.
}
\label{fig:parity_corr}
\end{figure*}

This asymmetry is physical as further shown in Fig.~\ref{fig:parity_corr}. The light
subhalos sit far outside their own truncation radii from the image ($b\gg
r_{\rm t}$), and thus, each acts essentially as a point mass.
The leading perturbation
to the image is therefore the tidal shear $T=\sum_i m_i/b_i^2$. The net (de)magnification
$\mathrm{dc}$ correlates with $T$ with \emph{opposite sign} for the two parities
(Spearman rank correlation $\simeq+0.51$ for the minimum and $\simeq-0.40$ for the
saddle).
The magnitude grows with the macro GO magnification $\sqrtmu$. The
opposite-sign correlation is the signature of the parity mechanism: a given tidal
perturbation magnifies a minimum and demagnifies a saddle. This is the WO
manifestation of the GO saddle fragility long known from quasar
flux-ratio anomalies~\cite{Schechter:2002dm,Kochanek:2003zc}: substructure
preferentially demagnifies saddle images. We find that the statement survives into
the per-image WO amplification factor, and that it lives in the
\emph{mean} (de)magnification while the fluctuation amplitude remains parity-blind.

\section{Detectability}
\label{sec:detect}

The per-image amplification factors are not directly observable.
The data carry
the \emph{total} lensed waveform $F_{\rm tot}(f) = F_{\rm min}(f)+F_{\rm sad}(f)$. In this section, we set out how the subhalo
imprint is extracted from that single observable and define the detection
statistic used in Sec.~\ref{sec:results}. Full derivations are in
Appendix~\ref{app:detect}.

\begin{figure*}[t]
\centering
\includegraphics[width=0.82\textwidth]{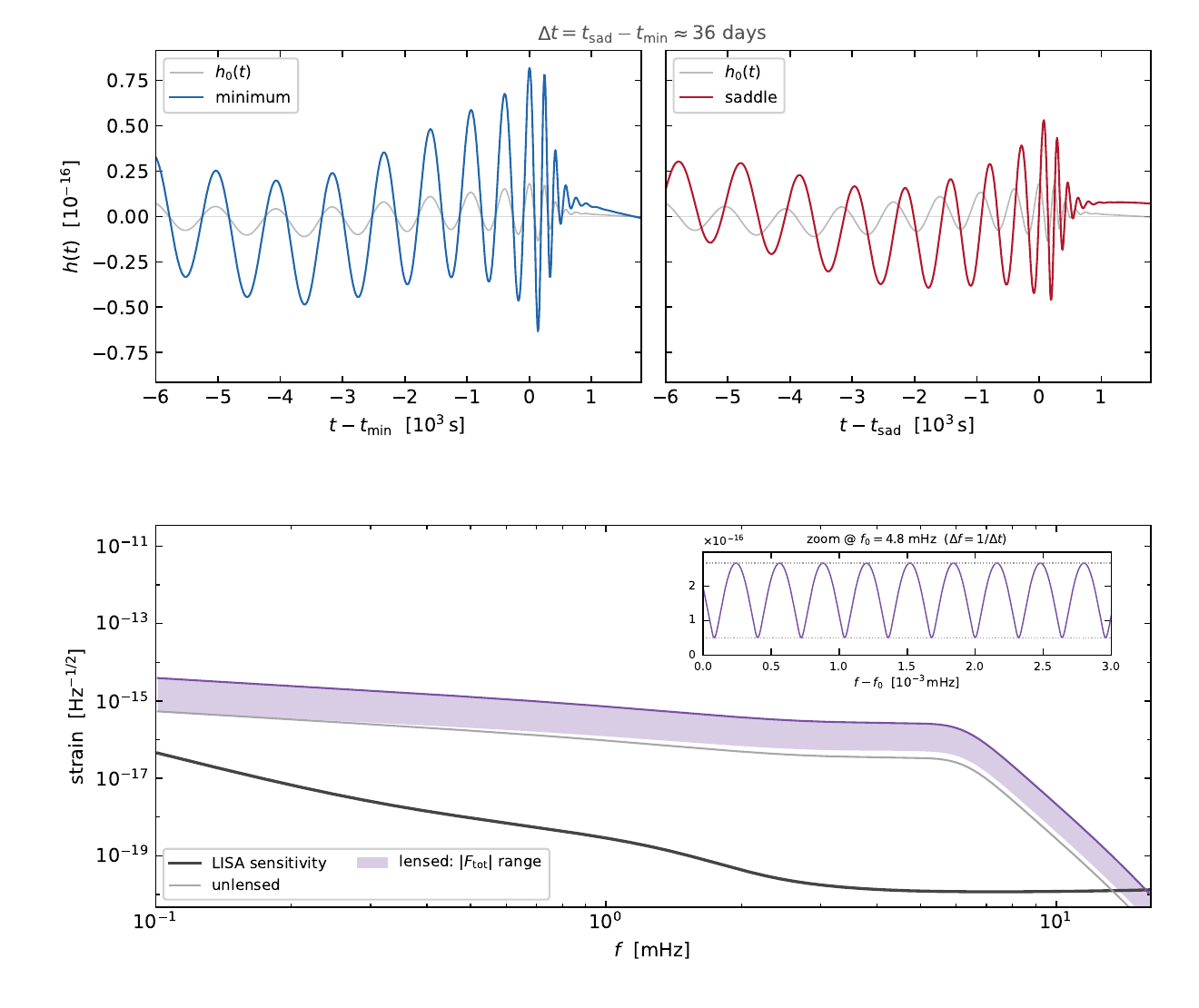}
\caption{How a dark matter substructure imprint enters a strongly lensed MBHB
signal, for a strongly imprinted realization [the extreme example~(c) of
Fig.~\ref{fig:topology}, at the $99.9$th percentile of the detection significance $\rho_{\rm sub}$].
\emph{Top:} time-domain strain of the two macro images about their respective
mergers. The signal-to-noise of an MBHB chirp accrues over only the last
$\sim$hours before coalescence, far shorter than the macro time delay
$\Delta t=t_{\rm sad}-t_{\rm min}$ (here $\approx36$~days, and weeks to a
year for a galaxy-scale lens). Therefore, the minimum (blue) and saddle (red) are received
as two cleanly separated chirps and can be analyzed
independently. \emph{Bottom:} the same signal as an amplitude spectral density
against the LISA sensitivity curve. The coherent two-image transfer function makes
$|F_{\rm tot}(f)|$ sweep between $|F_{\rm min}|+|F_{\rm sad}|$ and
$\big||F_{\rm min}|-|F_{\rm sad}|\big|$, filling the shaded band with a macro
interference fringe of spacing $1/\Delta t$ (inset). The WO subhalo
imprint is the slow, frequency-dependent modulation of the band \emph{envelope},
some four orders of magnitude coarser in $f$ than the macro fringe, such that the macro
delay and the per-image subhalo signal factorize. A frequency-independent
(de)magnification would merely rescale the envelope and is degenerate with the
per-image amplitude; only the envelope's frequency dependence carries the
substructure signal.}
\label{fig:buildup}
\end{figure*}

\subsection{The observable and its measurement}
\label{sec:obs}

The frequency-domain data are
\begin{equation}
d(f)=F_{\rm tot}(f)\,\tilde h(f;\bm\theta)+n(f),
\end{equation}
with $\tilde h$ the source waveform of parameters $\bm\theta$, $F_{\rm tot}$ the total
lens transfer function, and $n$ stationary Gaussian noise of one-sided power
spectral density $S_n$. Inference uses the matched-filter inner
product~\cite{Finn:1992wt,Cutler:1994ys}
\begin{equation}
\langle a|b\rangle=4\,\mathrm{Re}\!\int_0^\infty\frac{\tilde a\,\tilde b^{*}}
{S_n}\,\dd f,~~ \rho_0^{2}=\langle h|h\rangle,
\label{eq:inner}
\end{equation}
with $\rho_0$ the optimal S/N. Because the single-frequency variance of
$\hat F=d/\tilde h$ diverges,\footnote{With $\hat F=F_{\rm tot}+n/\tilde h$, the noise is
$\delta$-correlated in frequency, $\langle n(f)n^{*}(f')\rangle\propto S_n\,\delta(f-f')$,
so $\mathrm{Var}[\hat F(f)]\propto S_n\,\delta(0)/|\tilde h|^{2}\to\infty$.}
the transfer function is measured only band-averaged.
Since $F$ is smooth in $\ln f$, the natural resolution is one e-fold, and the
fractional precision on $F$ at frequency $f$ is
\begin{equation}
\frac{\sigma_F(f)}{|F|}\simeq\frac{1}{\rho_{\rm efold}(f)},~~
\rho_{\rm efold}^{2}(f)\equiv\frac{\dd\rho^{2}}{\dd\ln f}
=\frac{4\,|\tilde h(f)|^{2}f}{S_n(f)}.
\label{eq:efold}
\end{equation}
The noise on $F$, and hence on the subhalo modulation $\eta(f)$, is therefore
\emph{colored}: the imprint is measurable only where it overlaps the
S/N-carrying band, a fact that controls the detection significance below.

\subsection{Demodulating the macro images}
\label{sec:demod}

The data contain not the per-image amplification factors but their coherent sum. The two macro images arrive with a relative delay $\Delta t$ and
interfere,
\begin{equation}
F_{\rm tot}(f)=\sum_{j}\sqrt{\mu_{j}}\,\ee^{-\ii\pi n_j/2}\,
\big[1+\eta_j(f)\big]\,\ee^{2\pi\ii ft_j},
\label{eq:Ftot}
\end{equation}
with $n_j=0,1$ the Morse index of the minimum and the saddle and $\eta_j$ the per-image
WO modulation [Eq.~\eqref{eq:mdef}]. The two GO terms beat against
each other, modulating $|F_{\rm tot}(f)|$ with a rapid macro fringe of period
$1/\Delta t$ in frequency, on top of which the percent-level per-image subhalo
imprints $\eta_j$ ride. Extracting the $\eta_j$ from this dominant interference pattern is
the central measurement problem, and what makes it tractable is the wide separation
between the macro delay and the per-image time scales (Fig.~\ref{fig:buildup}). The
signal-to-noise of an MBHB chirp accrues over only the last $\sim$hours before
coalescence, far shorter than $\Delta t$. In the time domain, the two images are therefore
received as two well-separated chirps and may be analyzed separately.
Equivalently, in the frequency domain, since the macro fringe of spacing $1/\Delta t$ is far finer than
the e-fold scale over which the slowly varying $\eta_j$ evolve, the two factorize.

The same separation can be stated formally in the conjugate (cepstral) domain.
(This is not a necessary extra step beyond the time-domain windowing above, but
it makes the macro delay manifest as a blob spacing.) The Fourier
transform
\begin{equation}
g(t)=\int F_{\rm tot}(f)\,\ee^{-2\pi\ii ft}\,\dd f
\end{equation}
places the contribution of image $j$ in a compact blob centered on its arrival time
$t_j$, of width set by the per-image WO structure $t_{\rm sub}$, the
characteristic subhalo time delays of order $G m/c^{3}$, that is, a fraction of second to minutes
for the relevant $10^{4}$--$10^{7}\Msun$ perturbers. The macro delay
$\Delta t$, by contrast, is weeks to a year for a galaxy-scale lens. Therefore,
$t_{\rm sub}\ll\Delta t$ and the minimum and saddle blobs are cleanly resolved.
The two parities are thus separated not by any lens model, but by their different
arrival times: gating $g(t)$ about each $t_j$ and transforming back
(equivalently, windowing the two chirps apart directly in the time domain)
isolates the per-image transfer function $\sqrt{\mu_j}\,\ee^{-\ii\pi n_j/2}\,[1+\eta_j(f)]$.
In practice, however, one transforms the data $d=F_{\rm tot}\tilde h+n$, not $F_{\rm tot}$ itself.
Multiplying $F_{\rm tot}$ by the source $\tilde h(f)$ convolves each image with the compact chirp $h(t)$:
this leaves its center $t_j$ fixed and only broadens the blob to the chirp's S/N-accrual
duration (hours, still $\ll\Delta t$), leaving the separation argument intact. 

The GO prefactor $\sqrt{\mu_j}\,\ee^{-\ii\pi n_j/2}$ cannot be divided out: $\mu_j$ is
unknown and, together with the Morse phase and the arrival-time phase, is degenerate
with the source amplitude, coalescence phase, and merger time. What is recovered is
therefore $\sqrt{\mu_j}\,(1+\eta_j)$, on which the detection statistic of
Sec.~\ref{sec:stat} operates, projecting out that degenerate prefactor and retaining
only the non-absorbable frequency structure of $\eta_j$. The macro delay is delivered
for free as the blob spacing and needs no lens prior; only its absolute zero point is
degenerate with the coalescence time, which does not enter $\eta_j$.

At the high S/N of a strongly lensed event, the partition is lossless.
The deterministic finite-band sidelobes of each blob fall as $1/|t-t_j|$ and
are removed by a window taper, leaving the per-image measurement-noise floor
$\sigma_{\eta_j}(f)\simeq1/[\sqrt{\mu_j}\,\rho_{\rm efold}(f)]$
(Appendix~\ref{app:matched-filter}).
In our tests, the recovered $\sqrt{\mu_j}\,(1+\eta_j)$ reproduces the input to a fractional
accuracy of at most $\sim7\times10^{-7}$ across the $1000$ realizations, far below the
percent-level subhalo imprint. This is a deterministic, source-model-independent test
of the image separation itself; the construction, windowing, and the leakage it measures
are detailed in Appendix~\ref{app:demod}.
Thus, although the per-image
amplification factors $F_j$ are never measured in isolation, each per-image
transfer function $\sqrt{\mu_j}\,(1+\eta_j)$ is recoverable from the single observed
$F_{\rm tot}$, and the subhalo information it carries as the non-absorbable frequency
structure of $\eta_j$ is what the detection statistic below operates on.

\subsection{Detection statistic}
\label{sec:stat}

The subhalo contribution to the waveform of image $j$ is
$\delta h_{j}=\sqrt{\mu_j}\,\ee^{2\pi \ii f t_j-\ii\pi n_j/2}\,\eta_j\,\tilde h$. Not all of it is detectable: an overall
amplitude, a constant phase, and an arrival-time shift are degenerate with the
per-image lens extrinsic parameters $\{$amplitude, Morse phase, arrival time$\}$~\cite{Ezquiaga:2020gdt},
i.e., with the templates $\{1,\ \ii,\ 2\pi\ii f\}$. These are the leading
frequency-domain shapes of those three operations on the signal: $1$ is a constant rescaling
the amplitude, $\ii$ is a constant phase (associated to unknown coalescence phase $\phi_c$), and $2\pi\ii f$ is a
phase linear in $f$, i.e., an arrival-time shift (unknown coalescence time $t_c$).
We therefore retain only the
component of $\eta_j$ that cannot be absorbed by adjusting those three (its part
orthogonal, in the noise-weighted inner product below, to the span of $\{1,\ii,2\pi\ii f\}$),
\begin{equation}
\rho_{{\rm sub},j}=\sqrt{\mu_j}\,\big\lVert \eta_j^{\perp}\big\rVert_W,~~
\langle a,b\rangle_W=4\,\mathrm{Re}\!\int a\,b^{*}\,\frac{W(f)}{S_n(f)}\,\dd f,
\label{eq:rhosub}
\end{equation}
with the polarization-summed weight $W=|\tilde h_+|^{2}+|\tilde h_\times|^{2}$
and $\tilde h_+(f)$ and $\tilde h_\times(f)$ are the two frequency-domain polarization waveforms. Up to the
inclination-averaging factor of Sec.~\ref{sec:sources}, $W(f)$ equals the squared waveform
amplitude $|\tilde h(f)|^2$,
and
$\langle a,b\rangle_W$ is the noise-weighted inner product that induces the norm
$\lVert a\rVert_W=\langle a,a\rangle_W^{1/2}$ used here; in particular
$\lVert 1\rVert_W=\rho_0$.

Because the two images arrive at different macro delays,
they are statistically independent, and the combined significance adds in
quadrature,
\begin{equation}
\rho_{\rm sub,comb}=\sqrt{\rho_{{\rm sub},\rm min}^{2}+\rho_{{\rm sub},\rm sad}^{2}}\,.
\label{eq:rhocomb}
\end{equation}
A useful order-of-magnitude form is $\rhosub\approx\sqrtmu\,\rho_0\,
\varepsilon_{\rm eff}$, with $\varepsilon_{\rm eff}$ the $|\tilde h|^2/S_n$-weighted
rms of the non-absorbable frequency structure of $\eta$. An important consequence of
the projection is that the per-image \emph{mean} (de)magnification, the parity
asymmetry of Sec.~\ref{sec:results}, which enters $\eta$ as a frequency-independent
offset, is absorbed into the amplitude template and is \emph{not} detectable in a
single event; the single-event significance comes entirely from the
frequency-dependent ripple. The mean (de)magnification is nonetheless a real and
statistically robust population property, detectable in principle through its
correlation with the magnification across many events.

The complex modulation carries two physically distinct signals, and the statistic
separates them cleanly. Writing $\eta=\eta_{\rm a}+\ii\,\eta_{\rm p}$,
the real part $\eta_{\rm a}$ is the
amplitude modulation and the imaginary part $\eta_{\rm p}$ the phase modulation.
Both follow directly as the real and imaginary parts of the computed
$\eta_j=F_j/F_{{\rm GO},j}-1$, with $F_{{\rm GO},j}=\sqrt{\mu_j}\ee^{-\ii\pi n_j/2}$ the GO
prefactor, and need no separate fit. In the percent-level perturbative regime
$|\eta_j|\ll1$, and $\eta_{\rm a},\eta_{\rm p}$ then coincide with the amplitude and phase
modulations to first order. Because the real
template $1$ acts only on $\eta_{\rm a}$ (removing its band-averaged offset) while the
imaginary templates $\ii$ and $2\pi\ii f$ act only on $\eta_{\rm p}$ (removing a
constant and a linear-in-$f$ phase, i.e., $\phi_c$ and $t_c$), the cross terms in the
Gram matrix of $\{1,\ii,2\pi\ii f\}$ vanish and the two channels decouple in the
inner product of Eq.~\eqref{eq:rhosub}. The significance therefore splits exactly,
\begin{equation}
\rho_{{\rm sub},j}^{2}=\rho_{{\rm amp},j}^{2}+\rho_{{\rm phase},j}^{2},
\label{eq:ampphase}
\end{equation}
with the amplitude channel
$\rho_{{\rm amp},j}^{2}=\mu_j\lVert\eta_{{\rm a},j}-\langle\eta_{{\rm a},j}\rangle\rVert_W^{2}$,
its band-averaged offset $\langle\eta_{{\rm a},j}\rangle$ (the average of $\eta_{{\rm a},j}$ over
frequency weighted by $W/S_n$, the matched-filter weight, rather than a uniform mean) removed,
and the phase channel
$\rho_{{\rm phase},j}^{2}=\mu_j\lVert\eta_{{\rm p},j}-(\phi_c+2\pi t_c f)\rVert_W^{2}$,
its constant and linear-in-$f$ parts removed. The amplitude and phase imprints are
therefore separately accessible.
These two weighted subtractions are exactly the operation that forms $\eta^\perp$:
because the Gram matrix of $\{1,\ii,2\pi\ii f\}$ is block diagonal, the projection separates
into removing the $W$-weighted mean of $\eta_{\rm a}$ and the $W$-weighted constant and
linear-in-$f$ part of $\eta_{\rm p}$.

The amplitude channel dominates the combined detection, carrying $\simeq70$--$74\%$
of the power in $\rho^{2}$ across the three source masses, with the phase channel
carrying the remaining $\sim$quarter. Both channels are
thus substantial, and the
combined statistic uses both automatically. The slightly stronger suppression of the
phase channel follows from its projecting out one more template ($\phi_c$ and $t_c$)
than the amplitude channel (the offset alone).

We project out only the extrinsic set $\{$amplitude$,\,\phi_c,\,t_c\}$ and not the intrinsic
binary parameters (masses, spins), whose chirp-like frequency dependence is
distinct from the broadband WO ripple. Qualitatively, the WO modulation is broadband
and slowly varying in $\ln f$, whereas the derivatives of the waveform with respect to
the intrinsic parameters are sharply chirp-localized and oscillate on the much finer
GW-phase scale; their overlap with the projected modulation $\eta^{\perp}$ is therefore
small, and marginalizing them would lower $\rho_{\rm sub}$ only modestly.
The resulting significances are therefore a
mild upper bound, and a fully marginalized Fisher analysis is left to future work.

\subsection{Sources and noise}
\label{sec:sources}

We consider fiducial MBHB sources with mass ratio $q\equiv m_1/m_2=4$,
zero spin, at the
lens-fixed source redshift $z_S=1.5$ (luminosity distance of $11.2\unit{Gpc}$), and three total
source-frame masses $M_{\rm BH,tot}\in\{3\times10^5,\,10^6,\,3\times10^6\}\Msun$. The
waveforms are the aligned-spin \textsc{IMRPhenomD}
model~\cite{Husa:2015iqa,Khan:2015jqa} evaluated with
\textsc{LALSuite}/\textsc{PyCBC},
with detector-frame masses
$M_{\rm BH,tot}(1+z_S)$. The LISA sensitivity is the sky-averaged analytic curve
of Ref.~\cite{Robson:2018ifk} including the $4$-yr galactic confusion foreground, and
significances are quoted inclination-averaged (a factor $\sqrt{0.4}\simeq0.63$ on
the optimal S/N). The corresponding unlensed optimal S/N values are
$\rho_0=1.0\times10^{3}$, $2.3\times10^{3}$, and $1.8\times10^{3}$ for the three
masses~\cite{LISA:2017pwj}, peaking near $10^{6}\Msun$, where the innermost-stable-circular-orbit
frequency $f_{\rm ISCO}\sim1.8\unit{mHz}$ coincides with the LISA sensitivity
bucket.

\subsection{Detection significance}
\label{sec:res_detect}

\begin{figure*}[t]
\centering
\includegraphics[width=0.98\textwidth]{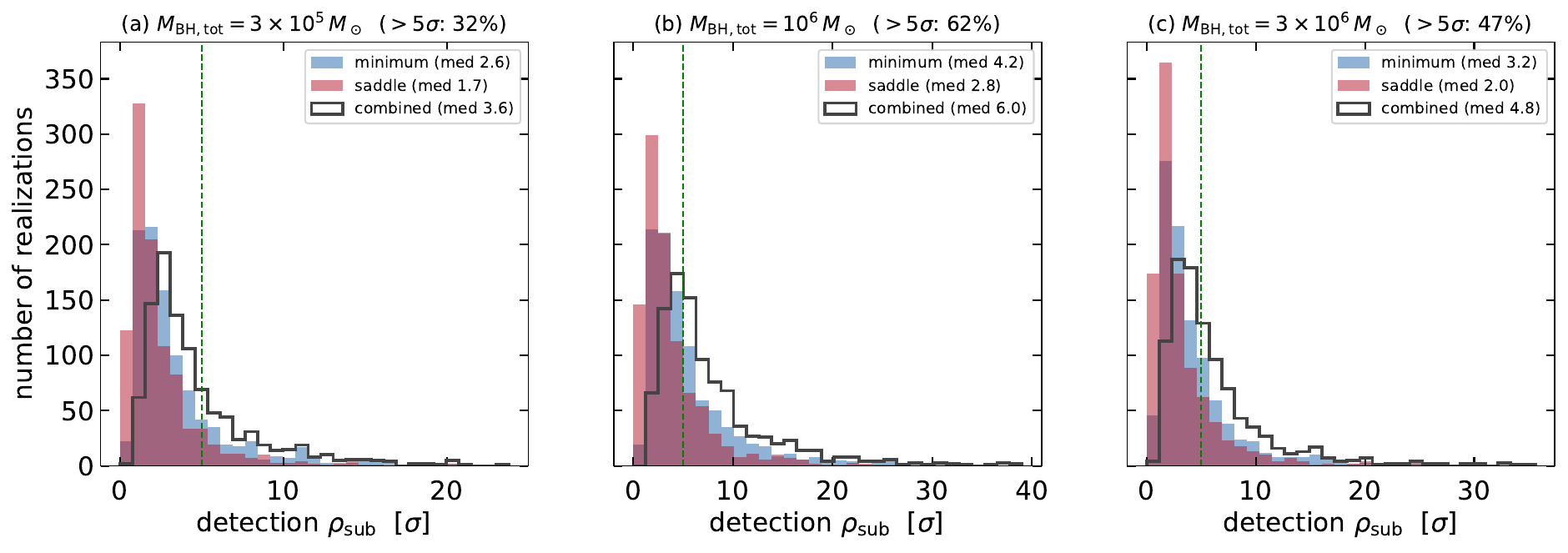}
\caption{Distribution of the subhalo detection significance $\rho_{{\rm sub}}$ over
the ensemble (inclination-averaged) for the three fiducial source masses:
(a)~$3\times10^{5}\Msun$, (b)~$10^{6}\Msun$, and (c)~$3\times10^{6}\Msun$ ($z_S=1.5$).
In each panel, the minimum (blue), saddle (red), and combined (black step) distributions
are shown; the dashed vertical line marks $5\sigma$. The combined detection exceeds
$5\sigma$ in $32\%$, $62\%$, and $47\%$ of realizations for the three masses,
respectively. The weak, strong, and extreme realizations of
Fig.~\ref{fig:topology} lie at the $5$th, $95$th, and $99.9$th percentiles of
$\rho_{\rm sub}$ (fiducial $10^{6}\Msun$ panel; $\approx2$, $22$, and
$103\,\sigma$).}
\label{fig:detect}
\end{figure*}

Figure~\ref{fig:detect} shows the distribution of the detection statistic
$\rho_{{\rm sub}}$ [Eqs.~\eqref{eq:rhosub}--\eqref{eq:rhocomb}] over the ensemble
for the three fiducial source masses ($3\times10^{5}$, $10^{6}$, and
$3\times10^{6}\Msun$), for the minimum, the saddle, and the combined
two-image signal.
The significances are summarized in
Table~\ref{tab:sigma}. For the fiducial $10^{6}\Msun$ source, near the peak of the
LISA MBHB sensitivity, the combined minimum-plus-saddle detection exceeds $5\sigma$
in the \emph{majority} ($62\%$) of subhalo realizations, with a median combined
significance $\rho_{\rm comb}\simeq6.0$. The lighter and heavier fiducial masses
give median $\rho_{\rm comb}\simeq3.6$ ($32\%$ above $5\sigma$) and $\simeq4.8$
($47\%$), respectively, tracking the unlensed S/N $\rho_0$ to which $\rho_{{\rm sub}}$
is proportional. Table~\ref{tab:sigma} resolves the detectability by image and
threshold: for the fiducial source, the minimum exceeds $3\sigma$ ($5\sigma$) in
$68\%$ ($40\%$) of realizations and the saddle in $48\%$ ($23\%$), while the combined
two-image signal reaches $89\%$ ($62\%$). The minimum is the more significant image,
both because it is more
magnified ($\sqrtmu_{\rm min}>\sqrtmu_{\rm sad}$) and because its non-absorbable
ripple is marginally larger. The saddle nonetheless contributes a comparable
significance, and including it raises the combined detectability above the
single-image value by the quadrature of Eq.~\eqref{eq:rhocomb}. Because
$\rho_{{\rm sub}}$ scales linearly with $\rho_0$, these numbers map directly onto any
individual loud LISA event through its own S/N.

The order-of-magnitude spread in $\rho_{\rm sub}$ across realizations is set by a
few subhalos at the smallest impact parameters. The imprint weighs perturbers as
$\sim m/b^2$ and peaks where a subhalo's Fermat delay falls in the sensitive band. Two
realizations with visually near-identical real-space subhalo maps can therefore differ
widely in detectability, as in the weak, strong, and extreme examples of
Fig.~\ref{fig:topology}.
A leave-one-out test bears this out: in the extreme example~(c), removing
the single closest massive subhalo ($m\approx6\times10^{6}\,M_\odot$ at impact parameter
$\approx0.007\,R_{\rm s}$, the bright point marked by a green circle in
Fig.~\ref{fig:topology}) lowers $\rho_{\rm sub}$ by $\sim90\%$.
The detectability therefore
cannot be read off the real-space configuration and must be computed from $F(w)$. 

\begin{table*}[t]
\caption{Detectability of the subhalo imprint over the Monte Carlo ensemble,
inclination-averaged, for the three fiducial MBHB source masses ($q=4$, $z_S=1.5$).
$\rho_0$ is the unlensed optimal S/N. The remaining columns give
the fraction of realizations whose subhalo detection significance $\rho_{\rm sub}$
[Eqs.~\eqref{eq:rhosub}--\eqref{eq:rhocomb}] exceeds $3\sigma$ and $5\sigma$, for the
minimum, the saddle, and the combined two-image signal.}
\label{tab:sigma}
\begin{ruledtabular}
\begin{tabular}{cccccccc}
 & & \multicolumn{3}{c}{$\rho_{\rm sub}>3\sigma$}
   & \multicolumn{3}{c}{$\rho_{\rm sub}>5\sigma$} \\
\cline{3-5}\cline{6-8}
$M_{\rm BH,tot}\,[\Msun]$ & $\rho_0$ & minimum & saddle & combined
 & minimum & saddle & combined \\
\colrule
$3\times10^{5}$ & $1.0\times10^{3}$ & $40\%$ & $24\%$ & $61\%$
 & $20\%$ & $10\%$ & $32\%$ \\
$1\times10^{6}$ & $2.3\times10^{3}$ & $68\%$ & $48\%$ & $89\%$
 & $40\%$ & $23\%$ & $62\%$ \\
$3\times10^{6}$ & $1.8\times10^{3}$ & $53\%$ & $34\%$ & $77\%$
 & $29\%$ & $17\%$ & $47\%$ \\
\end{tabular}
\end{ruledtabular}
\end{table*}

\subsection{Convergence and robustness}
\label{sec:res_robust}

The signal is dominated by subhalos in the immediate neighborhood of the image, which the ensemble samples within $\Rnear$. We test convergence
against this choice with an independent ensemble that doubles the sampling radius
to $2\Rnear$ (a fourfold increase in subhalo count): the two agree within
the realization scatter. For the fiducial $10^{6}\Msun$ source the median combined
detection significance changes from $\rho_{\rm comb}=5.97$ ($<\Rnear$) to $6.10$
($<2\Rnear$) and the fraction above $5\sigma$ is $62\%$ in both, with the lighter and
heavier masses agreeing to within $\lesssim3\%$ in the median and $\lesssim3$
percentage points in the threshold fractions; the parity statistics are likewise
unchanged. The signal is therefore converged at $<\Rnear$.

\subsection{Dependence on source position}
\label{sec:res_ysrc}

\begin{figure*}[t]
\centering
\includegraphics[width=0.9\textwidth]{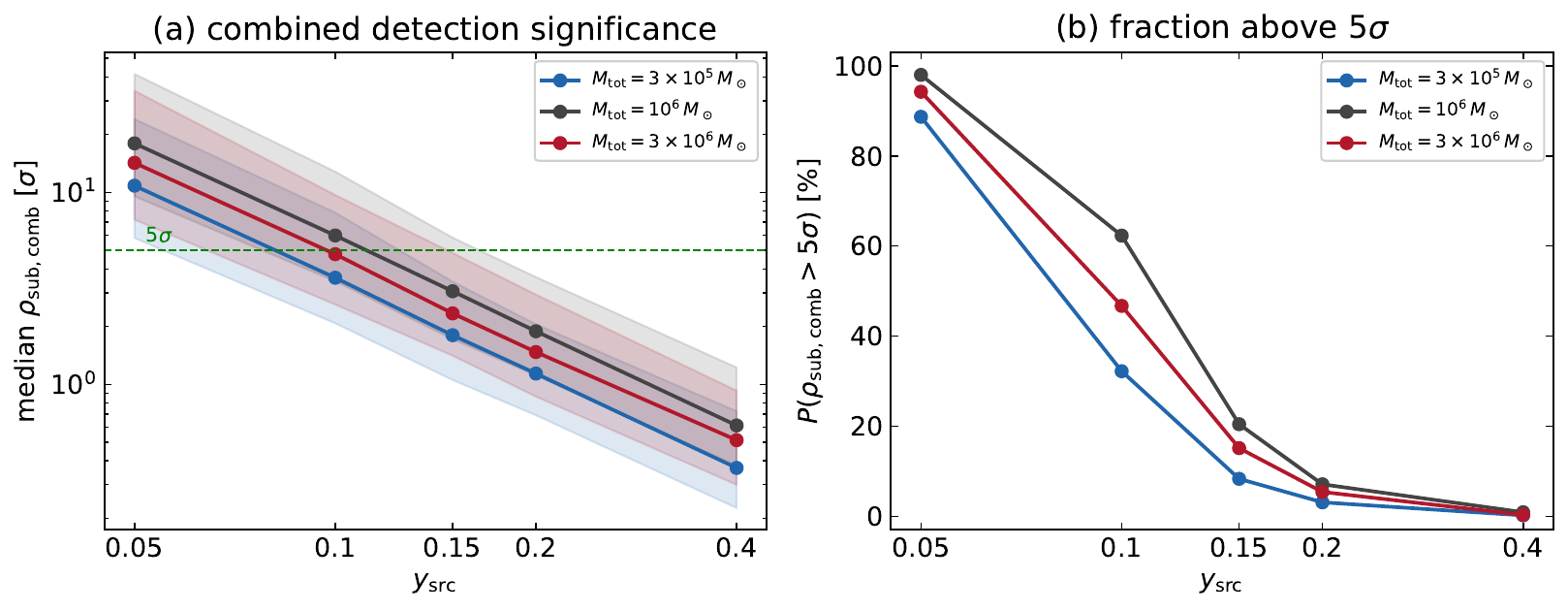}
\caption{Detectability versus source impact parameter $y_{\rm src}$, for the three
fiducial source masses ($1000$ realizations per point, inclination-averaged).
(a)~Median combined significance $\rho_{\rm sub,comb}$ with the $16$--$84\%$ band;
the dashed horizontal line marks $5\sigma$. (b)~Fraction of realizations above $5\sigma$. The
significance falls steeply off the caustic, $\rho_{\rm comb}\propto
y_{\rm src}^{-1.6}$, such that a robust detection requires $y_{\rm src}\lesssim0.1$.}
\label{fig:ysrc}
\end{figure*}

The detection significance is set by the macro magnification through
$\rho_{\rm sub}\propto\sqrtmu$ [Eq.~\eqref{eq:rhosub}], and $\mu$ rises steeply as
the source approaches the lens caustic. The fiducial choice $y_{\rm src}=0.1$ is
not representative of all configurations. To map this dependence we repeat the full
calculation for source impact parameters
$y_{\rm src}\in\{0.05,\,0.1,\,0.15,\,0.2,\,0.4\}$, with $1000$ realizations each and
the lens and subhalo model held fixed. Figure~\ref{fig:ysrc} shows the median
combined significance and the fraction above $5\sigma$ as functions of
$y_{\rm src}$. The significance falls steeply and nearly as a power law,
$\rho_{\rm comb}\propto y_{\rm src}^{-1.6}$, dropping by a factor $\simeq3$ for each
doubling of $y_{\rm src}$. For the fiducial $10^{6}\Msun$ source the median combined
significance crosses $5\sigma$ near $y_{\rm src}\simeq0.11$. Therefore, a robust per-event
detection requires the strongly magnified regime $y_{\rm src}\lesssim0.1$. For
$y_{\rm src}\gtrsim0.2$, the imprint reaches $5\sigma$ in only a few percent of
realizations. The fiducial configuration therefore sits at the edge of the
favorable region, and the steep dependence implies that a LISA strong-lensing
substructure search is dominated by the most highly magnified, near-caustic events,
an essential input for any population forecast (Sec.~\ref{sec:discussion}).

\section{Discussion}
\label{sec:discussion}

The relevance of these signatures rests on the existence of strongly lensed MBHB
events for LISA. Recent forecasts span of order $0.1$ to a few hundred strongly
lensed MBHB mergers over a multi-year LISA
mission~\cite{Sereno:2010dr,Oguri:2018muv,Gutierrez:2025ymd}, the wide range
reflecting the uncertain high-redshift massive-black-hole population (in
particular, the light- versus heavy-seed scenarios), together with the high-redshift
lens population, so that the absolute rate carries an order-of-magnitude
uncertainty that is beyond the scope of this work to refine.
Such an event is moreover straightforward to \emph{identify} as a strong
lens~\cite{Haris:2018vmn,Hannuksela:2019kle}, since it appears as two near-identical chirps separated by the macro time
delay. Therefore, the
practical limitation is the intrinsic occurrence rate rather than the recognition of
the event. Such
events are rare, but each is loud, since the same large magnification that makes the
configuration a strong lens also supplies the high S/N that the
WO measurement requires. A single such event near a macro critical curve is,
therefore, a sensitive substructure probe, and our fiducial configuration, a modest
source offset behind a galaxy-scale lens, is representative rather than fine-tuned.

Combining these factors gives a rough sense of the mission yield: the most
optimistic forecast of $\sim200$ lensed MBHBs, the $\approx15\%$ of detectable
events that fall within $y_{\rm src}\lesssim0.1$ of the caustic once magnification
bias is included (Paper~I), and a combined $5\sigma$ detection fraction of
$\sim30$--$62\%$ across the fiducial source masses (Table~\ref{tab:sigma}), give $10$--$20$
events carrying a detectable substructure imprint.
These detection fractions are for the fiducial lens and source redshifts
($z_L=0.5$, $z_S=1.5$). We have not surveyed their configuration dependence, and the yield
here should be read as representative of the fiducial geometry rather than a
redshift-averaged rate, with the redshift folding left to the population forecast below.
Conservative source and lens populations reduce this to well
below one. The estimate is thus dominated by the order-of-magnitude uncertainty in
the strong-lensing rate, and is best read as a per-event sensitivity that a single
loud event already realizes.

Because the signal is set by the compactness of the perturbers and by the macro
amplification, it is enhanced for more concentrated substructure than the truncated
CDM subhalos considered here. Scenarios that produce denser low-mass
structure, such as a population of primordial black
holes~\cite{Sasaki:2018dmp,Ando:2024ghr}, gravothermal core collapse in
self-interacting dark matter~\cite{Tulin:2017ara,Ando:2024kpk}, or the
massive $\rho \propto r^{-1.5}$ prompt cusps in warm dark matter~\cite{Delos:2025ees}, would raise the
modulation amplitude and hence the detectability.
The WO signal thus offers a route to constrain not
only the abundance but also the internal density of dark substructure at
$10^{4}$--$10^{7}\,M_\odot$ scales, below those accessible to electromagnetic probes.

Two further directions remain.
The amplitude of the saddle modulation carries a residual method dependence at the
sub-percent level that an independent direct diffraction computation could remove.
A population forecast that folds the per-event significances of
Table~\ref{tab:sigma} through the strong-lensing MBHB rate and the distribution of
source positions and redshifts would convert our per-event results into an expected
number of substructure detections over the LISA mission.
We defer these to future work, together with the application to
substructure beyond the CDM paradigm.

\section{Conclusions}
\label{sec:conclusions}

The nature of dark matter on subgalactic scales remains an open question, and the
abundance and compactness of substructure in the $10^{4}$--$10^{7}\,M_\odot$ range,
below the reach of electromagnetic probes, is one of its sharpest discriminants.
We have extended the wave-optics treatment of dark matter substructure in strongly
lensed gravitational waves from the minimum image of Paper~I to the saddle-point
image, and we have quantified the detectability of the combined signal with LISA.
Evaluating the amplification factor at a saddle is numerically delicate, because the
equal-arrival-time contours are open and the subhalo signal is a small difference of
large terms. We presented a time-domain method, built from an analytic
common-template subtraction and an analytic far-field tail, and validated it
against an independent area integral of the arrival-time surface. Both this saddle
calculation and the minimum-image calculation of Paper~I are implemented in a single public
code, \textsc{Sazanami}.
A by-product of the derivation is a clean physical statement of why the saddle differs
from the minimum. Because the saddle's equal-arrival-time contour is open and unbounded, the
perturbing potential is sampled over a logarithmically growing arc; the subhalo imprint then
decays more slowly with delay, carrying an extra logarithmic enhancement, than at the
minimum, whose closed, bounded contour leaves no such term.

Across a Monte Carlo ensemble of cold dark matter subhalo realizations, subhalos
generically imprint percent-level amplitude and $\sim10^{-2}\,$rad phase modulations
on both image parities, and the mean (de)magnification splits by parity, with the
minimum net magnified and the saddle net demagnified, correlating with the local
tidal field with opposite sign. This is the wave-optics counterpart of the
geometric-optics saddle fragility, and it resides in the frequency-independent (mean)
part of the modulation while the fluctuating amplitude is parity-blind.
Treating the total
lensed waveform as the only observable, we demodulated the macro images to recover
the per-image modulations and constructed a matched-filter statistic that retains
only the non-absorbable frequency structure. For a fiducial $10^{6}\,M_\odot$
massive-black-hole-binary source whose position lies in the strongly magnified
region near the lens caustic, the combined minimum-plus-saddle detection exceeds
$5\sigma$ in the majority of realizations. A survey of the source impact parameter $y_{\rm src}$
shows that this significance falls steeply off the caustic,
$\rho_{\rm sub}\propto y_{\rm src}^{-1.6}$. A robust per-event detection
requires $y_{\rm src}\lesssim0.1$, and the most highly magnified events dominate any
search. Folding these per-event significances through current strong-lensing rate
forecasts, the most optimistic source and lens populations give $10$--$20$ substructure detections over a multi-year LISA mission, while conservative
populations give well below one; the result is therefore most robustly stated per
event, where even a single loud lensed event is informative.

Strongly lensed gravitational waves are therefore a sensitive,
complementary probe of dark matter substructure at $10^{4}$--$10^{7}\,M_\odot$
scales inaccessible to electromagnetic observations. Because the signal grows with
the compactness of the perturbers, it is a particularly promising diagnostic of
scenarios that produce denser low-mass structure, such as primordial black holes or
core-collapsed self-interacting dark matter, for which the fragile saddle is the
most responsive image. A population forecast over the LISA strong-lensing rate, a
fully marginalized detection analysis, and the application to alternative dark
matter models are natural next steps, and would turn the parity asymmetry into a
quantitative constraint on the abundance and compactness of dark substructure.

\begin{acknowledgments}
This work was supported by JSPS KAKENHI Grant Number 24K07039. The wave-optics amplification factors were computed with the
\textsc{GLoW} package~\cite{Villarrubia-Rojo:2024xcj}, the subhalo populations
were generated with \textsc{SASHIMI}~\cite{Hiroshima:2018kfv}, and the source
waveforms and LISA sensitivity were evaluated with \textsc{LALSuite}~\cite{lalsuite}/\textsc{PyCBC}~\cite{pycbc}
and \textsc{astropy}~\cite{astropy}.
\end{acknowledgments}

\appendix
\section{Saddle-point numerics}
\label{app:saddle}

This appendix collects the technical details of the saddle-point calculation of
Sec.~\ref{sec:saddle}: the patch-bounded quadratic-saddle template, the numerical
tracing of the open contour and its area-method validation, the derivation of the
far-field tail and its Fourier transform, and the removal of the residual
numerical fringes left by the Filon transform.

\subsection{The patch-bounded quadratic-saddle template}
\label{app:quadtemplate}

The template $I_{\rm quad}$ of Eq.~\eqref{eq:Iquad} is the co-area
integral~\eqref{eq:Itau} of the bare local quadratic. In the Hessian eigenframe,
$\phi-\phi_{\rm sad}=\lambda_+x_+^2/2-|\lambda_-|x_-^2/2$, the
co-area integrand reduces to a constant in the contour's natural parameter
$\zeta$,
\begin{equation}
\frac{\dd\ell}{|\nabla\phi|}=\sqrtmu\,\dd\zeta ,
\label{eq:coarea_const}
\end{equation}
the \emph{same} for both image types; only the range of $\zeta$ and its
trigonometric vs.\ hyperbolic character differ. For a minimum the iso-arrival
contour is the closed ellipse
$x_\pm=\sqrt{2s/\lambda_\pm}\,(\cos\zeta,\sin\zeta)$, $\zeta\in[0,2\pi)$, giving the constant $I_{\rm GO}=2\pi\sqrtmu$. For a saddle ($s>0$) it is the open
hyperbola $x_+=\sqrt{2s/\lambda_+}\cosh\zeta$,
$x_-=\sqrt{2s/|\lambda_-|}\sinh\zeta$, and the arc must be cut where the quadratic
ceases to hold, at the circular patch $x_+^2+x_-^2\le R_{\rm c}^2$ that the
tracer itself imposes. Eliminating $x_-^2=(\lambda_+x_+^2-2s)/|\lambda_-|$ at the
boundary fixes the exit point and hence $\zeta_{\max}$,
\begin{equation}
\begin{aligned}
x_{+,\rm patch}^2&=\frac{|\lambda_-|R_{\rm c}^2+2s}{\lambda_++|\lambda_-|},\\[2pt]
u_{\max}=\cosh\zeta_{\max}&=\sqrt{x_{+,\rm patch}^2\,\lambda_+/(2s)}\,.
\end{aligned}
\label{eq:umax_app}
\end{equation}
Each branch contributes $\sqrtmu$ integrated over $|\zeta|\le\zeta_{\max}$ (both
vertex halves), and the two branches give
$I_{\rm quad}=4\sqrtmu\,\mathrm{arccosh}(u_{\max})$, i.e.\ Eq.~\eqref{eq:Iquad};
the $s<0$ branch follows from $\lambda_+\leftrightarrow|\lambda_-|$. As
$s\to0$, $\mathrm{arccosh}(u_{\max})\to-(\ln s)/2+\mathrm{const}$ and
$I_{\rm quad}\to-2\sqrtmu\ln|s|$, the saddle geometric-optics logarithm. This is
the finite-region template of Ref.~\cite{Tambalo:2022plm}, with their constant
contour limit $\delta\tau$ resolved here into the explicit, weakly $s$-dependent
circular patch, $\delta\tau=\lambda_+x_{+,\rm patch}^2/2$. Because the
tracer cuts $I_{\rm full}$ at the same radius $R_{\rm c}$, for a bare quadratic
macro $\delta I=I_{\rm full}-I_{\rm quad}$ vanishes identically; the residual that
enters Eq.~\eqref{eq:recipe} therefore carries only the subhalo imprint.

\subsection{Tracing the open saddle contour}
\label{app:trace}

The open saddle contour is traced as four half-arcs (two hyperbola branches, two
vertex halves). Each half-arc is integrated as an ordinary differential equation
along the contour, in the co-area parameter $\sigma$ with $\dd I/\dd\sigma=R$, by an
adaptive, error-controlled Runge--Kutta integrator (\textsc{GSL}'s RK8(7), absolute and
relative tolerance $10^{-10}$) that follows the iso-arrival level set
$\{\phi=\tau\}$ outward from its waist until the contour reaches the patch radius
$R_{\rm c}$. There is thus no fixed arc-length grid; $\Delta\sigma_0$ below is the
integrator's initial step in $\sigma$.
With too coarse an
arc step the tracer intermittently drops one or two half-arcs at scattered $s$,
leaving $I_{\rm full}$ short by a multiple of $I_{\rm quad}/4$ and producing spurious
spikes in the envelope; crucially the affected points are returned as nominally
converged, and a failure-only retry does not catch them. The control parameter is the
step $\Delta\sigma_0=R_{\rm c}/n_{\rm div}$, not $n_{\rm div}$ itself: holding
$n_{\rm div}$ fixed gives a coarser step for larger patches and reintroduces the
dropouts. Fixing the step instead ($\Delta\sigma_0=0.01$ in patch units, capped
slightly finer for the largest patches) makes the tracer reach $R_{\rm c}$ on every
half-arc. The dropouts thus disappear at their source and the spurious spikes vanish. The
few points the escape-bounded tracer genuinely cannot complete are flagged unconverged and
repaired by filling: linear interpolation of $I$, or pinning the residual to zero in the
immediate near-saddle region where it vanishes analytically.

We confirm the corrected trace against an arbiter that uses no contour tracing at
all: $I(s)$ is also the derivative of the area of the sub-level set of the
arrival-time surface,
\begin{equation}
I(s)=\frac{\dd}{\dd s}\,\mathrm{Area}\{\bm{x}:\ \phi(\bm{x})-\phi_{\rm sad}\le s\},
\label{eq:area}
\end{equation}
evaluated by cell-counting on the arrival-time surface at two grid resolutions. This
lens-agnostic ground truth reproduces our traced $I(s)$ to a median fractional
difference $\sim10^{-3}$ across the band, confirming the arrival-time integral that
enters Eq.~\eqref{eq:recipe}; the same area method validated the minimum in Paper~I.

\subsection{Far-field tail: co-area derivation and Fourier transform}
\label{app:tail}

Write the full arrival-time surface as the quadratic saddle plus the subhalo
potential, $\phi=\phi_{\rm quad}+\psi$. To first order the co-area integral
[Eq.~\eqref{eq:Itau}] is perturbed by
\begin{equation}
\delta I(s)=-\frac{\dd}{\dd s}\oint_{\phi_{\rm quad}=s}
\frac{\psi\,\dd\ell}{|\nabla\phi_{\rm quad}|}
\equiv-\frac{\dd}{\dd s}\langle\psi\rangle_s .
\end{equation}
For the open saddle the weight $\dd\ell/|\nabla\phi_{\rm quad}|$ is the
$s$-independent constant $g_\infty=1/\sqrt{\lambda_+|\lambda_-|}=\sqrtmu$
[Eq.~\eqref{eq:coarea_const}], and the arc runs to
$\zeta_{\max}=\kappa-\tfrac12\ln s$ (hence $\cosh\zeta_{\max}\propto1/\sqrt s$),
its logarithmic extent. With the truncated halo's monopole
$\psi\simeq M_{\rm t}\ln r$ and $\ln r=\ln(2s)/2+\ln\Lambda(\zeta)/2$
(so $r\propto\sqrt s$ along the arc), $\langle\psi\rangle_s=\sqrtmu M_{\rm t}\,
\mathcal{G}(s)$ with $\mathcal{G}$ geometric and quadratic in $\ln s$.
Differentiating,
\begin{equation}
\delta I(s)=-\frac{\dd}{\dd s}\langle\psi\rangle_s=\frac{a+b\ln s}{s},
\qquad a,\,b\propto\sqrtmu\,M_{\rm t},
\label{eq:tail_app}
\end{equation}
i.e.\ Eq.~\eqref{eq:tail}: explicitly $b=\sqrtmu M_{\rm t}$ and
$a=\sqrtmu M_{\rm t}(\ln2+\gamma)$ with
$\gamma=\ln[(\lambda_+^{-1}+|\lambda_-|^{-1})/4]$, $M_{\rm t}$ the dimensionless
mass defined there. The $R_{\rm c}$-dependent constant $\kappa$ cancels between
the two pieces of $\mathcal{G}$ (each $\propto(\ln s)^2$), leaving $a$ and $b$ independent
of the patch radius.
\emph{Both} come from the monopole. For a subhalo population, it is the mass
\emph{enclosed within the window}, $M_{\rm enc}$ (the subhalos with
$|s_i|<s_{\rm hi}$), which is realization-dependent but carries no free parameter:
the positions enter only at dipole order ($\sim(\ln s)/s^{3/2}$,
faster-decaying), and the amplitude follows from the enclosed subhalo masses alone. The $\ln s$ enhancement is the product of the potential's $\ln r$
and the open contour's $\ln s$ extent. A minimum, closed and bounded, lacks the
second logarithm and its residual falls as a pure $1/s$.

\paragraph*{Fourier transform.}
The Filon quadrature of Sec.~\ref{sec:saddle_method} transforms $\delta I$ over the
finite window $|s|\le s_{\rm hi}$, the matching point at which the traced residual
is handed to the analytic tail. We adopt $s_{\rm hi}=6000$ (units of
$s=\tau-\phi_{\rm sad}$), placed beyond the light-subhalo Fermat delays $s_k$ so the
residual has settled onto its $1/|s|$ plateau, yet inside the quadratic patch where
the open arc is reliable (the patch radius is in fact enlarged from $s_{\rm hi}$,
to $R_{\rm c}\ge\sqrt{2s_{\rm hi}/(0.3\,|\lambda_-|)}$). The $s$-grid is a
two-sided geometric grid over $|s|\le s_{\rm hi}$ ($45$ nodes per branch). Beyond it
the residual is continued analytically as the tail $\widetilde{T}$
[Eq.~\eqref{eq:Ttilde}], whose kernels follow from elementary integrals: on the
$s>0$ branch, with $u=ws$ and $x=w\,s_{\rm hi}$,
\begin{equation}
\int_{s_{\rm hi}}^{\infty}\frac{\ee^{\ii ws}}{s}\,\dd s
=-\mathrm{Ci}(x)+\ii\left[\frac{\pi}{2}-\mathrm{Si}(x)\right]\equiv J(x),
\end{equation}
the $s<0$ branch its conjugate $\overline{J}$.\footnote{$\mathrm{Ci}(x)=
-\int_x^{\infty}(\cos t/t)\,\dd t$, $\mathrm{Si}(x)=\int_0^{x}(\sin t/t)\,\dd t$.}
The $\ln|s|/|s|$ part gives $K$ of Eq.~\eqref{eq:JK}, which has no elementary closed
form. With $z=-\ii w$, it equals
$\partial_a\big[z^{-a}\Gamma(a,z\,s_{\rm hi})\big]_{a=0}$ ($\Gamma$ the upper
incomplete gamma), evaluated to machine precision. The amplitudes $C_\pm,D_\pm$
[Eq.~\eqref{eq:tailC}] are the per-branch $a,b$ for $M_{\rm enc}^\pm$ (the subhalos
with $0<\pm s_i<s_{\rm hi}$), computed from the in-window subhalo masses with no
fit; a least-squares fit on the plateau $[0.3\,s_{\rm hi},s_{\rm hi}]$ reproduces
them to $\sim\!10^{-4}$ in the envelope as a cross-check. The two finite-window
fringes (one at the Filon/tail joint, one from the coarse transform grid) are
removed as described in Appendix~\ref{app:fringe}.

\subsection{Numerical fringes from the Filon transform}
\label{app:fringe}

The residual $\delta I$ is transformed in two pieces: a Filon quadrature over the
window $|s|\le s_{\rm hi}$ and the analytic tail $\widetilde{T}$ beyond it
(App.~\ref{app:tail}). Handled naively, each piece imprints a fringe on the
envelope that is common to all realizations (hence not physical subhalo
signal), and we remove both at negligible cost. The physical fringes, at the
subhalo Fermat delays $s_k=|\bm{x}_s^{\mathsf T}A\bm{x}_s|/2$, are untouched.

\paragraph{Filon/tail joint ring.} The analytic tail
$(C_\pm+D_\pm\ln|s|)/|s|$ (App.~\ref{app:tail}), computed from the enclosed
monopole, does not match the traced residual exactly at the window edge
$|s|=s_{\rm hi}$. Splicing it onto the Filon window there leaves a step
\begin{equation}
\Delta_\pm=\delta I(\pm s_{\rm hi})-\big(C_\pm+D_\pm\ln s_{\rm hi}\big)/s_{\rm hi}
\end{equation}
between the traced and the analytically-continued residual, whose Fourier
transform is a constant-amplitude ring at conjugate delay $s_{\rm hi}$. We cancel
it by adding the (Abel-regularised) transform of that step, one per branch, to the
tail,
\begin{equation}
\widetilde{T}_{C^0}(w)=-\frac{\Delta_+\,\ee^{\ii w s_{\rm hi}}}{\ii w}
+\frac{\Delta_-\,\ee^{-\ii w s_{\rm hi}}}{\ii w},
\end{equation}
which restores continuity of the spliced residual across the joint and removes the
ring.

\paragraph{Filon grid-resolution fringe.} The traced residual is smooth and well
resolved by the coarse production $s$-grid ($45$ nodes per branch), but the Filon
rule itself under-resolves the oscillatory kernel $\ee^{\ii ws}$ between the
sparse large-$|s|$ nodes once $w\,\Delta s\gtrsim\pi$, leaving a grid-tied fringe
that grows toward high $w$. Because the trace is already converged, we cure this
by refining only the transform: $\delta I$ is interpolated with a monotone
piecewise-cubic Hermite scheme (PCHIP), which fits a cubic through each pair of samples
with slopes chosen to preserve the data's monotonicity and avoid the overshoot of an
ordinary cubic spline, per branch onto a fine geometric $s$-grid ($2000$ nodes per side)
before the Filon rule is applied.
This changes neither the trace nor its cost and is re-derivable
from the stored coarse $I(s)$.

With both corrections the saddle envelope is clean across the band.
Left uncorrected, the two fringes are small but clearly visible. Across our realizations the
Filon/tail joint ring contributes a band-uniform ripple of $\sim0.1$--$0.15\%$ in the
envelope, and the grid-resolution fringe a further $\sim0.02$--$0.08\%$ that grows toward
high frequency, together $\lesssim0.2\%$. This is comparable to the smallest physical
subhalo imprints and is enough to roughen the otherwise smooth saddle envelope, which is why
we remove both.

\section{Detection formalism}
\label{app:detect}

\subsection{Matched-filter identities and the per-image noise floor}
\label{app:matched-filter}

We collect the matched-filter identities underlying Sec.~\ref{sec:detect}. With
the one-sided power spectral density $S_n$ defined by
$\langle\tilde n(f)\tilde n^*(f')\rangle=S_n(f)\delta(f-f')/2$ and the
reality condition $\tilde n(-f)=\tilde n^*(f)$, the likelihood-natural two-sided
inner product reduces, for real signals, to the one-sided form of
Eq.~\eqref{eq:inner}: the factor $4=2\times2$ combines the fold of negative
frequencies with the one-sided convention, and the $\mathrm{Re}$ is the imprint of
the conjugate negative-frequency half.

For the matched-filter statistic $x=\langle d|h\rangle$ with $d=h+n$,
\begin{equation}
\mathbb{E}[x]=\langle h|h\rangle,~~
\mathrm{Var}[x]=\mathrm{Var}\langle n|h\rangle=\langle h|h\rangle,
\end{equation}
the second equality following because the cross terms $\propto\delta(f+f')$ have no
support on $f,f'>0$. Hence the optimal signal-to-noise ratio is
$\rho_0=\sqrt{\langle h|h\rangle}$, and more generally
$\mathrm{Cov}[\langle n|a\rangle,\langle n|b\rangle]=\langle a|b\rangle$, the
basis of the Fisher matrix used to project out the lens and source parameters in
Eq.~\eqref{eq:rhosub}.

The transfer function is recovered only as a band average. The single-frequency
estimator $\hat F=d/\tilde h$ has variance $\propto\delta(0)$; over a band the
amplitude estimator $\hat F_0=\langle d|\tilde h\rangle_{\rm band}/
\langle\tilde h|\tilde h\rangle_{\rm band}$ has
$\mathrm{Var}[\hat F_0]=\langle\tilde h|\tilde h\rangle_{\rm band}^{-1}$, so
that with one e-fold of bandwidth $\sigma_F/|F|\simeq1/\rho_{\rm efold}$ with
$\rho_{\rm efold}^2=4|\tilde h|^2 f/S_n$ as in Eq.~\eqref{eq:efold}. The per-image
measurement-noise floor that the lossless demodulation of Sec.~\ref{sec:demod}
leaves behind, $\sigma_{\eta_j}(f)\simeq1/[\sqrtmu_j\,\rho_{\rm efold}(f)]$, has an
inverse-square integral over $\ln f$ that reproduces the per-image significance
$\rho_{{\rm sub},j}$ of Eq.~\eqref{eq:rhosub}.

The LISA sensitivity~\cite{Robson:2018ifk} used throughout is the sky-averaged
one-sided strain power spectral density
\begin{equation}
\begin{aligned}
S_n(f)={}&\frac{10}{3L^2}\left[P_{\rm OMS}
+\frac{2\,(1+\cos^2(f/f_*))\,P_{\rm acc}}{(2\pi f)^4}\right]\\
&\times\left(1+\frac{6}{10}\,\frac{f^2}{f_*^2}\right)+S_c(f),
\end{aligned}
\end{equation}
with arm length $L=2.5\times10^{9}\unit{m}$, $f_*=c/(2\pi L)=19.1\unit{mHz}$,
optical-metrology and acceleration noises $P_{\rm OMS}$, $P_{\rm acc}$, and the
$4$-yr galactic confusion foreground $S_c$. We adopt the coefficients of
Ref.~\cite{Robson:2018ifk}.

\subsection{Validating the image separation: the demodulation test}
\label{app:demod}

The claim of Sec.~\ref{sec:demod} that the two macro images separate losslessly is
verified by a direct round-trip on $F_{\rm tot}$ that is deterministic and
source-model-independent: no waveform, sensitivity curve, or noise enters. For each of the
$1000$ realizations we take the per-image factors
$\sqrt{\mu_j}\,\ee^{-\ii\pi n_j/2}[1+\eta_j(f)]$, and synthesize the coherent sum of Eq.~\eqref{eq:Ftot} with
symmetric arrival times $t_{\rm min}=-\Delta t/2$, $t_{\rm sad}=+\Delta t/2$ for the fiducial
macro delay $\Delta t\simeq36\unit{d}$. The sum is sampled on $f\in[10^{-4},10^{-1}]\unit{Hz}$ with a
step fine enough to oversample the macro fringe $1/\Delta t$ ($\gtrsim5$ samples per fringe)
and apodized by a Tukey band taper (raised-cosine roll-off over the outer $10\%$ of each band
edge), which suppresses the finite-band sidelobes $\propto1/|t-t_j|$.

The Fourier dual, $$g(t)=\int F_{\rm tot}(f)\,\ee^{-2\pi\ii ft}\,\dd f,$$ evaluated by FFT,
places each image in a compact blob at its $t_j$ of width $\sim t_{\rm sub}(\ll\Delta t)$. A
smooth raised-cosine gate of half-width $0.4\,\Delta t$ centered on each $t_j$ isolates one
image (the two gates do not overlap). Inverse-transforming and removing the carrier
$\ee^{-2\pi\ii ft_j}$ returns the per-image transfer function $\sqrt{\mu_j}\,(1+\eta_j)$. We
compare this reconstruction to the (tapered) input over the flat interior of the band as the
maximum fractional deviation $\max_f[|F_{\rm rec}-F_{\rm in}|/|F_{\rm in}|]$, and report the
worst value over all realizations.

The worst-case fractional error is $\sim7\times10^{-7}$ at $\Delta t=36\unit{d}$. The residual
inter-image leakage is thus orders of magnitude below both the percent-level imprint $\eta_j$
and the per-image measurement-noise floor
$\sigma_{\eta_j}\simeq1/(\sqrt{\mu_j}\,\rho_{\rm efold})$: the demodulation introduces no
significant systematic. Note, however, this figure is the idealized limit of the cepstral test, in which
each blob has its intrinsic per-image width $t_{\rm sub}$. With the realistic chirp time scale
(Sec.~\ref{sec:demod}) the blobs broaden to the chirp's S/N-accrual duration and the
inter-image leakage grows. Since $\Delta t$ still vastly exceeds that duration, the leakage is
expected to remain well below the percent-level imprint.

\section{Subhalo population and lens model}
\label{app:population}

\subsection{Subhalo sampling}

Heavy ($m>10^9\Msun$) and light ($10^2$--$10^9\Msun$) subhalos are drawn from the
semi-analytic \textsc{SASHIMI} model~\cite{Hiroshima:2018kfv,Ando:2019xlm,
Ando:2020yyk}, with the three-dimensional number density
$n_{\rm sub}(r)\propto(r^2+R_{\rm s}^2)^{-3/2}$. Light subhalos near an image are
sampled in a projected disc of radius $\Rnear$ with the expected
count fixed by the local cylinder weight
\begin{equation}
w_{\rm cyl}=w_{\rm sub}\,\frac{\pi R_{\rm sample}^2\,I_z}{\mathcal{V}_{\rm halo}},
\quad
I_z=\frac{2z_{\max}}{a^2\sqrt{z_{\max}^2+a^2}},
\end{equation}
with $a^2=R_{\rm img}^2+R_{\rm s}^2$, $z_{\max}=\sqrt{R_{\rm vir}^2-R_{\rm img}^2}$,
and $\mathcal{V}_{\rm halo}$ the virial normalization. Positions are
uniform in the disc, $r=\sqrt{u}\,R_{\rm near}$. An independent ensemble within the radius of $2\Rnear$ serves as the convergence test of
Sec.~\ref{sec:res_robust}.

\subsection{Truncated-NFW lens and mass match}

Each subhalo is the $n=1$ truncated-NFW profile of Eq.~\eqref{eq:tnfw}, whose
enclosed mass within radius $c\,r_{\rm s}$, in units of $4\pi\rho_{\rm s}
r_{\rm s}^3$, is
\begin{equation}
I_1(c)=\frac{c^2}{(c^2+1)^2}\big[(c^2-1)\ln c+\pi c-(c^2+1)\big].
\end{equation}
\textsc{SASHIMI} fixes the truncation assuming an abrupt three-dimensional cut,
$I_\infty(c)=\ln(1+c)-c/(1+c)$. To preserve the bound mass exactly under the
profile change we solve
\begin{equation}
I_1\!\big(c^{(1)}\big)=I_\infty\!\big(c^{(\infty)}\big)
\end{equation}
for the $n=1$ concentration $c^{(1)}$, which fixes $r_{\rm t}^{(1)}=c^{(1)}r_{\rm s}$;
the match is exact to machine precision. We use the closed-form truncated-NFW deflection,
potential, and convergence of Ref.~\cite{Baltz:2007vq} (equivalently the
\textsc{lenstronomy}~\cite{Birrer:2018xgm} implementation), with the off-center
lens obtained by rigid translation.

\subsection{External-field decomposition}

The potential is split as $\psi=\psi_{\rm macro}+\delta\psi$. The macro field
$\psi_{\rm macro}$ collects the host, galaxy, and heavy subhalos together with the
light subhalos that the thresholds of Sec.~\ref{sec:subs} classify as
GO, while $\delta\psi$ retains only the WO light subhalos.
Because the GO light subhalos perturb the arrival-time surface, we re-solve for the
stationary points of $\psi_{\rm macro}$ once they are included, verifying that the
image multiplicity is unchanged. We then build the WO frame at the
re-evaluated image. About that image the macro field is expanded to second order,
the Hessian
\begin{equation}
A=I-\nabla\nabla\psi_{\rm macro}\big|_{\bm{x}_{\rm img}}
=\begin{pmatrix}1-\kappa-\gamma_1 & -\gamma_2\\ -\gamma_2 & 1-\kappa+\gamma_1
\end{pmatrix},
\end{equation}
in terms of the local convergence $\kappa$ and shear $(\gamma_1,\gamma_2)$. This
guarantees that the macro field is exact to second order, the GO
stationary point is preserved, and there is no double counting between the macro
field and the explicit subhalo term in Eq.~\eqref{eq:Fw}.

\bibliography{refs}

\end{document}